\documentclass[fleqn,usenatbib]{mnras}

\usepackage{newtxtext,newtxmath}

\usepackage[T1]{fontenc}
\usepackage{ae,aecompl}
\usepackage{booktabs}
\usepackage{threeparttable}

\usepackage{float}

\usepackage{graphicx}	
\usepackage{amsmath}	
\usepackage{amssymb}	

\newcommand{\tessstarnameone}{TOI-150}
\newcommand{\tessstarnametwo}{TOI-163}
\newcommand{\starnameone}{TYC9191-519-1}
\newcommand{\starnametwo}{HD271181}
\newcommand{\codename}{\texttt{juliet}}

\title[2 \textit{TESS} Hot Jupiter Discoveries]{TOI-150b and TOI-163b: two transiting hot Jupiters, one eccentric and one inflated, revealed by \textit{TESS} near and at the edge of the JWST CVZ}
\author[D. Kossakowski et al.]{Diana Kossakowski,$^{1}$\thanks{E-mail: kossakowski@mpia.de}
N\'estor Espinoza,$^{1}$\thanks{Bernoulli Fellow}\thanks{IAU-Gruber Fellow}
Rafael Brahm,$^{2,3,4}$
Andr\'es Jord\'an,$^{3,4}$
\newauthor
Thomas Henning,$^{1}$
Felipe Rojas,$^{3,4}$
Martin K\"urster,$^{1}$
Paula Sarkis,$^{1}$
Martin Schlecker,$^{1}$
\newauthor
Francisco J. Pozuelos,$^{5,6}$
Khalid Barkaoui,$^{6,7}$
Emmanu\"el Jehin,$^{5}$
Micha\"el Gillon,$^{6}$
\newauthor
Elisabeth Matthews,$^{8}$
Elliott P. Horch,$^{9}$
David R. Ciardi,$^{10}$
Ian J. M. Crossfield,$^{8}$
\newauthor
Erica Gonzales,$^{11}$
Steve B. Howell,$^{12}$
Rachel Matson,$^{12}$
Joshua Schlieder,$^{13}$
\newauthor
Jon Jenkins,$^{12}$
George Ricker,$^{14}$
Sara Seager,$^{14,15,16}$
Joshua N. Winn,$^{17}$
Jie Li,$^{18}$
\newauthor
Mark E. Rose,$^{19}$
Jeffrey C. Smith,$^{18}$
Scott Dynes,$^{14}$
Ed Morgan,$^{8}$
\newauthor
Jesus Noel Villasenor,$^{14}$
David Charbonneau,$^{20}$
Tess Jaffe,$^{21,22}$
Liang Yu,$^{14}$
\newauthor
Gaspar Bakos,$^{17}$
Waqas Bhatti,$^{17}$
Fran\c{c}ois Bouchy,$^{23}$
Karen A. Collins,$^{20}$
\newauthor
Kevin I. Collins,$^{24}$
Zoltan Csubry,$^{17}$
Phil Evans,$^{25}$
Eric L. N. Jensen,$^{26}$
\newauthor
Christophe Lovis,$^{23}$
Maxime Marmier,$^{23}$
Louise D. Nielsen,$^{23}$
David Osip,$^{27}$
\newauthor
Francesco Pepe,$^{23}$
Howard M. Relles,$^{20}$
Damien S\'egransan,$^{23}$
Avi Shporer,$^{14}$
\newauthor
Chris Stockdale,$^{28}$
Vincent Suc,$^{3}$
Oliver Turner,$^{23}$
and St\'ephane Udry$^{23}$
\\
A list of the authors' affiliations can be found in Appendix \ref{affiliations}.
}

\date{Accepted XXX. Received YYY; in original form ZZZ}

\pubyear{2019}

\begin{document}
\label{firstpage}
\pagerange{\pageref{firstpage}--\pageref{lastpage}}
\maketitle

\begin{abstract}
We present the discovery of \starnameone b (\tessstarnameone b, TIC 271893367) and \starnametwo b (\tessstarnametwo b, TIC 179317684), two hot Jupiters initially detected using 30-minute cadence Transiting Exoplanet Survey Satellite \textit{(TESS)} photometry from Sector 1 and thoroughly characterized through follow-up photometry (CHAT, Hazelwood, LCO/CTIO, El Sauce, TRAPPIST-S), high-resolution spectroscopy (FEROS, CORALIE) and speckle imaging (Gemini/DSSI), confirming  the planetary nature of the two signals. 
A simultaneous joint fit of photometry and radial velocity using a new fitting package \codename\ reveals that \tessstarnameone b is a $1.254\pm0.016\ \textnormal{R}_\textnormal{J}$, massive ($2.61^{+0.19}_{-0.12}\ \textnormal{M}_\textnormal{J}$) hot Jupiter in a $5.857$-day orbit, while \tessstarnametwo b is an inflated ($R_\textnormal{P}$ = $1.478^{+0.022}_{-0.029} R_\textnormal{J}$, $M_\textnormal{P}$ = $1.219\pm0.11 \textnormal{M}_\textnormal{J}$) hot Jupiter on a $P$ = $4.231$-day orbit; both planets orbit F-type stars. A particularly interesting result is that \tessstarnameone b shows an eccentric orbit ($e=0.262^{+0.045}_{-0.037}$), which is quite uncommon among hot Jupiters. We estimate that this is consistent, however, with the circularization timescale which is slightly larger than the age of the system. These two hot Jupiters are both prime candidates for further characterization --- in particular, both are excellent candidates for determining spin-orbit alignments via the Rossiter-McLaughlin (RM) effect and for characterizing atmospheric thermal structures using secondary eclipse observations considering they are both located closely to the James Webb Space Telescope (JWST) Continuous Viewing Zone (CVZ). 

\end{abstract}




\section{Introduction}
We are now entering an exciting era with NASA's Transiting Exoplanet Survey Satellite \textit{(TESS)} mission \citep{ricker_tess:2016}, a nearly all-sky survey with the primary goal of uncovering and more so characterizing planets smaller than Neptune ($\lesssim 4 R_\oplus$) around nearby and bright stars ($V<13$). The expected yield for the short 2-minute cadence targets ($\sim$200,000) is roughly 1250 new transiting planets of various sizes \citep{barclay_tessyield:2018}, adding onto the already impressive quantity of $\sim$4,000 transiting planets discovered\footnote{As of March 11, 2019: https://exoplanetarchive.ipac.caltech.edu/} 
to date --- most of which come from the \textit{Kepler} \citep{borucki_kepler:2010} transit survey. 
The quantity of new discoveries can be imagined to be even higher when we include the longer 30-minute cadence targets, potentially increasing the yield to 25,000 \citep{barclay_tessyield:2018}. The opportunity for new world discoveries and classification is high considering that \textit{TESS} is focusing on the brightest neighboring stars, making it easier for ground-based instruments to follow-up the transit planet detections allowing for further, more detailed characterization. 

Among the diversity of new worlds to be discovered by \textit{TESS}, hot Jupiters --- planets of 
similar mass to Jupiter ($0.3M_J \lesssim M \lesssim 2M_J$) and with periods $P<10$ days \citep{dawson:2018} --- are naturally the most accessible to detect due to their size (relatively larger flux dip in light curve) and short orbiting periods (multiple transits for a given light curve time baseline). Their massive nature also makes them ideal targets for radial velocity (RV) follow-up, as this imposes large modulations in their host star's motion. 
\textit{TESS}, for this reason, will then be able to detect \textit{most} of the transiting hot Jupiters in our stellar neighborhood; HD 202772Ab \citep{wanghj:2019} and HD2685 b \citep{hd2685} are thus just the first of many to be detected by the mission.

Hot Jupiters are interesting objects on their own right, as they are objects that are still not well understood. For example, it is known that their radii are larger than expected from models of irradiated exoplanets \citep[see, e.g.,][and references therein]{TF:2018} --- however, the mechanism of this so-called ``radius inflation" is still not known. Their formation is also a mystery --- how giant exoplanets like these end up in short period orbits around their stars is still an open question in the field \citep[see][for a review]{dawson:2018}. A larger sample of exoplanets might help resolve these issues or help find new predictions for models to make --- for example, using the current sample of hot Jupiters, \cite{sestovic:2018} recently showed that the radius inflation might depend on mass. Using a similar sample, \cite{TF:2018} suggested that the efficiency with which energy is deposited in the interior of hot Jupiters to make them look inflated might depend on equilibrium temperature. \cite{Bailey:2018} recently showed that the period-mass 
distribution of hot Jupiters could be explained by in-situ formation of hot Jupiters. It is clear from studies like these that enlarging the sample of known, well-characterized hot Jupiters can aid in understanding their nature and evolution, and thus is an important endeavor to undertake.

In this work, we introduce the discovery and characterization of two new hot Jupiters, \tessstarnameone b and \tessstarnametwo b, whose signals were initially detected by \textit{TESS} long-cadence photometry and then thoroughly followed up by other photometric (CHAT, Hazelwood, LCO/CTIO, El Sauce, TRAPPIST-S) and spectroscopic ones (FEROS, CORALIE) ground-based facilities. 

The paper is structured as follows. In Section \ref{sec:data}, we present all of the photometric, spectroscopic, and speckle image observations gathered for both targets. In Section \ref{sec:analysis}, we focus on the characterization of the star and the planets in details using a joint analysis of the 
data, combining transit photometry and radial velocities. In Section \ref{sec:discussion}, we present a discussion on these targets and their qualifications as follow-up candidates for atmospheric characterization and spin-orbit alignment.

During the writing of this manuscript, another paper \citep{canas_toi150:2019} introduced the discovery of \tessstarnameone b. Though the paper delivered the planetary detection, we provide and present a more complete and thorough analysis with 4 photometric follow-up instruments and a total of 23 radial velocities (20 from FEROS and 3 from CORALIE), which in turn provides a precise constraint on the planetary 
and orbital parameters of the system. The inclusion of these extra radial velocity measurements, 
allow us to find a strong signal of an eccentric orbit for this exoplanet --- this is further discussed 
in Section \ref{subsubsec:eccentricity}.

\section{Data} \label{sec:data}
The photometric and high-resolution imaging observations were obtained as part of the \textit{TESS} Follow-up Program (TFOP)\footnote{https://tess.mit.edu/followup/}. All follow-up photometric data along with the speckle images were acquired via Exoplanet Follow-up Observing Program for TESS (ExoFOP-TESS). The radial velocities are presented in Table \ref{tab:rvdata}. We used the {\tt TESS Transit Finder}, which is a customized version of the {\tt Tapir} software package \citep{Jensen:2013}, to schedule photometric time-series follow-up observations. In addition, we worked with the AstroImageJ software package \citep{Collins:2017} to perform aperture photometry for most of these follow-up photometric observations, excluding CHAT which uses a separate pipeline (Jordan et al. in prep.). 
For \tessstarnameone, we have 5 photometric datasets (\textit{TESS}, LCO z and i bands, El Sauce, and TRAPPIST-S) and 2 radial velocity instruments (FEROS, CORALIE). The data alongside with the best model fits are plotted in Figures \ref{fig:toi150_phot} and \ref{fig:toi150_rvs}. For \tessstarnametwo\ we also have 5 photometric datasets (\textit{TESS}, CHAT, Hazelwood, LCO i band, and El Sauce) and 1 radial velocity instrument (FEROS). The data and model fits can be found in Figures \ref{fig:toi163_phot} and \ref{fig:toi163_rvs} --- these are 
detailed below. 

\subsection{\textit{TESS} Photometry} \label{subsec:TESSphot}
\textit{TESS} was designed to observe 26 $24^\circ \times 90^\circ$ sections of the sky (or ``sectors" --- 13 in the Northern and 13 in the Southern hemisphere), for which each is roughly observed for one 
month ($\sim$27 days) over the course of the planned two-year 
mission\footnote{https://tess.mit.edu/observations/}. The photometric bandpass of \textit{TESS} (600$-$1000 nm) is very similar to the $G_{rp}$ band pass  (630$-$1050 nm) for the $Gaia$ survey \citep[Data Release 2 (DR2)][]{gaiadr2summary}, a fact that will prove to be useful when looking for possible contaminating sources in the TESS photometry. Both targets, \starnameone\ (TIC 271893367, \tessstarnameone, \textit{Gaia} DR2 5262709709389254528) and \starnametwo\ (TIC 179317684, \tessstarnametwo, \textit{Gaia} DR2 51366259202463104), were observed in Sector 1 (from 2018 July 25 -- August 22) with the 30-minute cadence full-frame images (FFIs). Calibrated FFIs are conveniently available for quick download via the Mikulski Archive for Space Telescopes (MAST)\footnote{https://archive.stsci.edu/tess/; https://mast.stsci.edu/tesscut/} where the entire \textit{TESS} Input Catalog (TIC) is uploaded and where the archival lightcurve data produced by the Science Processing Operations Center (SPOC) pipeline reside \citep{jenkins_ffi:2016}. The lightcurves used for this work were taken from the \textit{TESS} alerts page, from which we extracted the Simple Aperture Photometry 
fluxes (\texttt{SAPFLUX}).

\begin{figure*}
\centering
\begin{minipage}{\textwidth}
  \includegraphics[width=\linewidth]{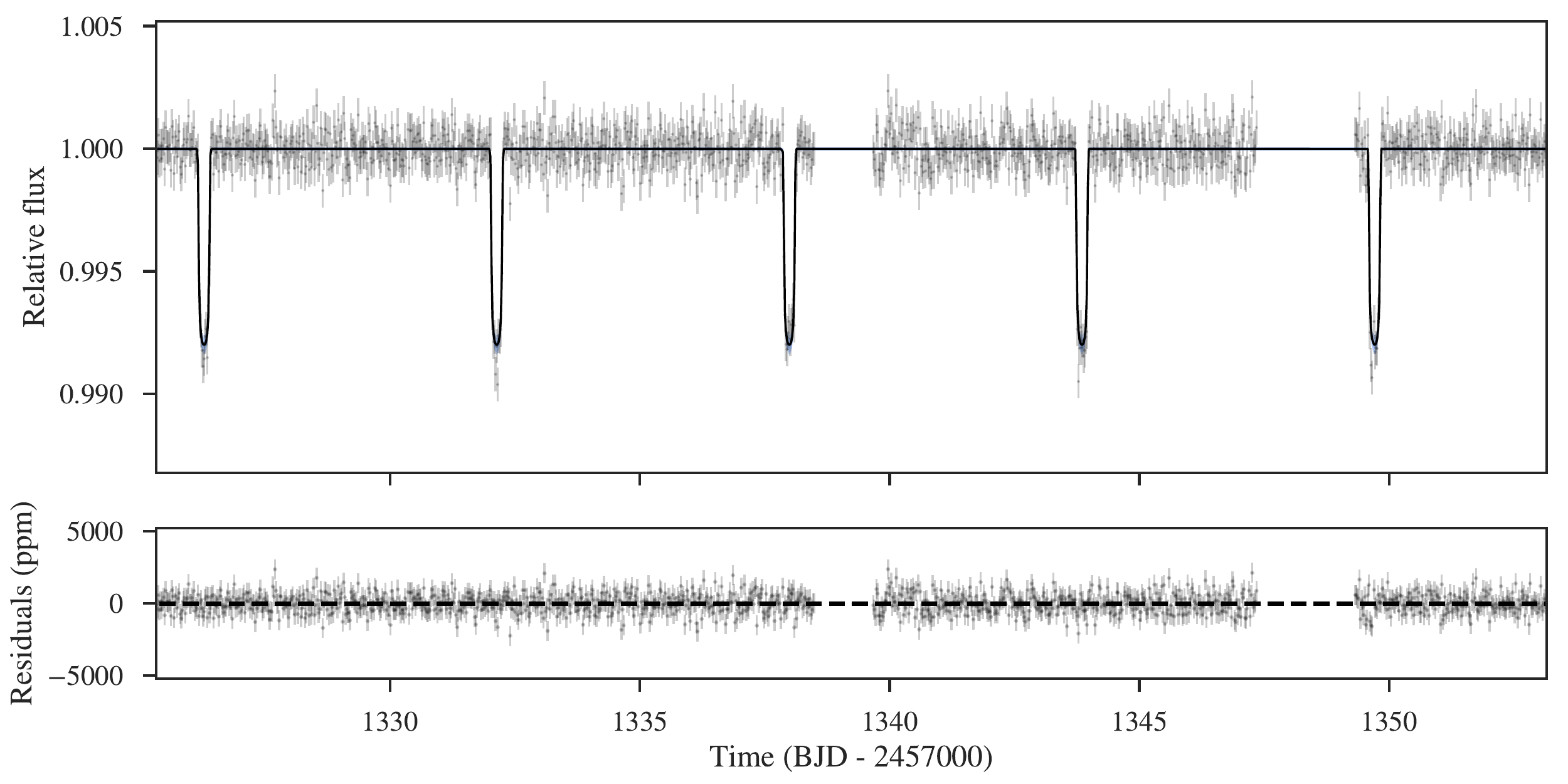}
\end{minipage}
\begin{minipage}{.33\textwidth}
  \centering
  \includegraphics[width=1\linewidth]{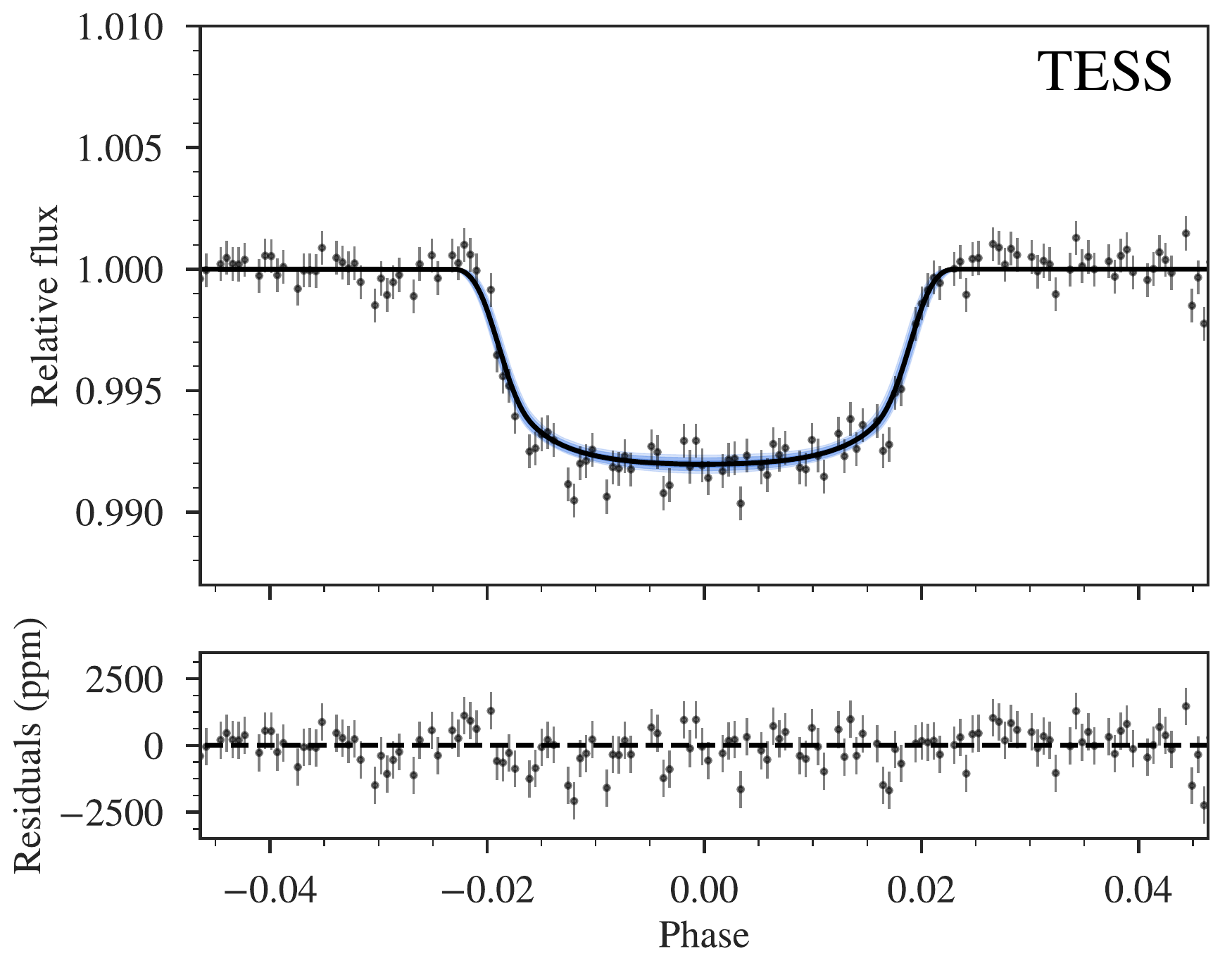}
\end{minipage}%
\begin{minipage}{.33\textwidth}
  \centering
  \includegraphics[width=1\linewidth]{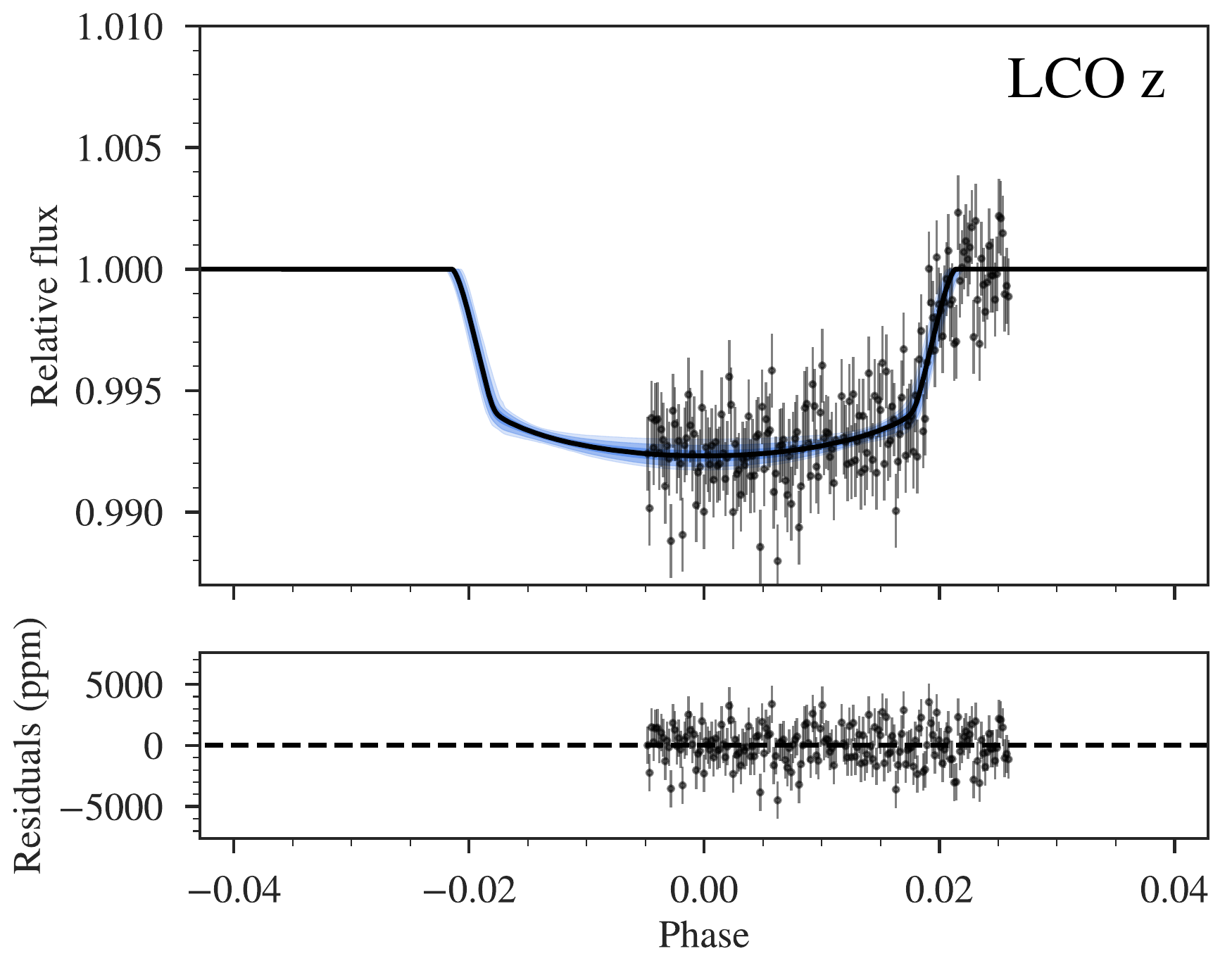}
\end{minipage}%
\begin{minipage}{.33\textwidth}
  \centering
  \includegraphics[width=1\linewidth]{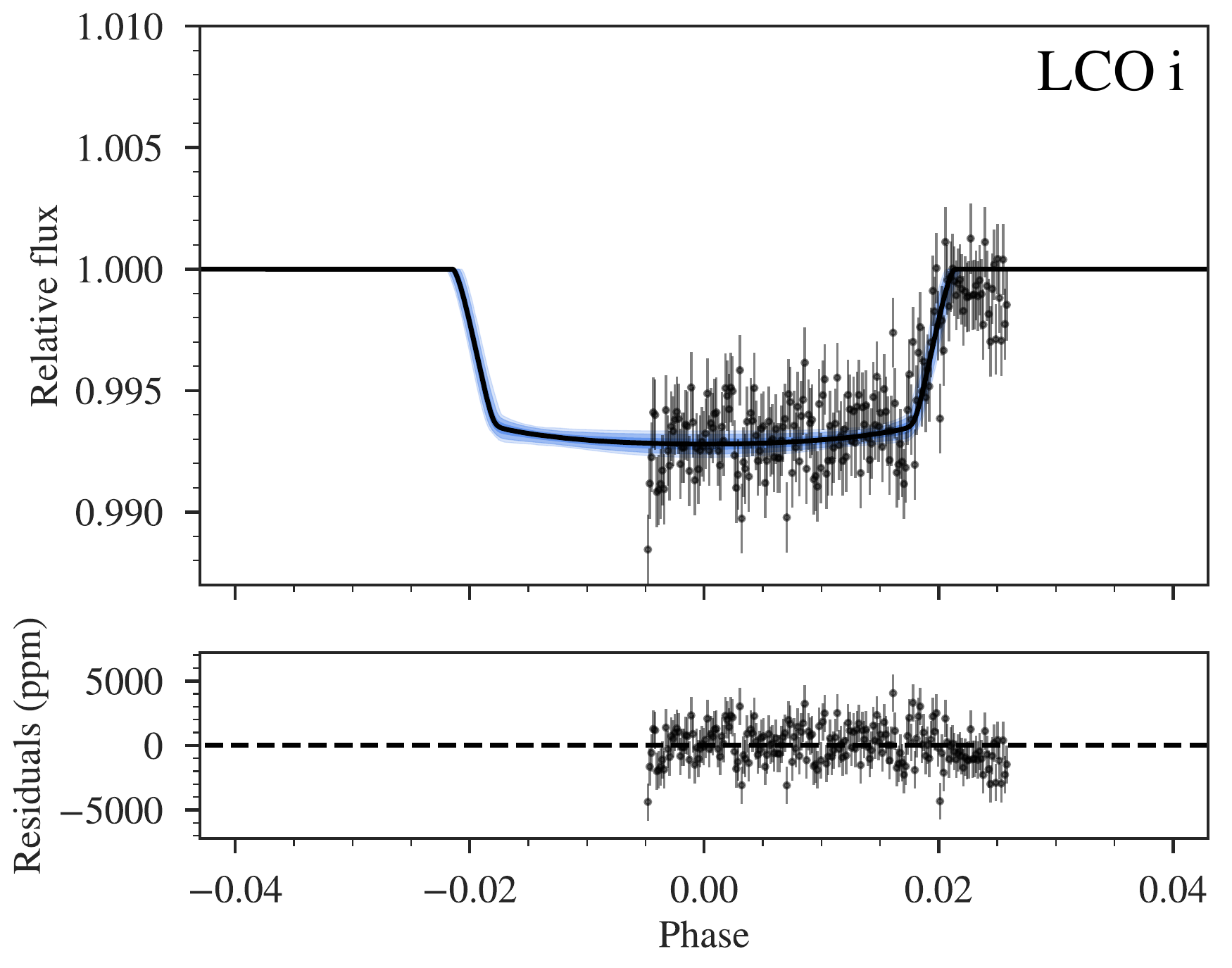}
\end{minipage}
\begin{minipage}{.33\textwidth}
  \centering
  \includegraphics[width=1\linewidth]{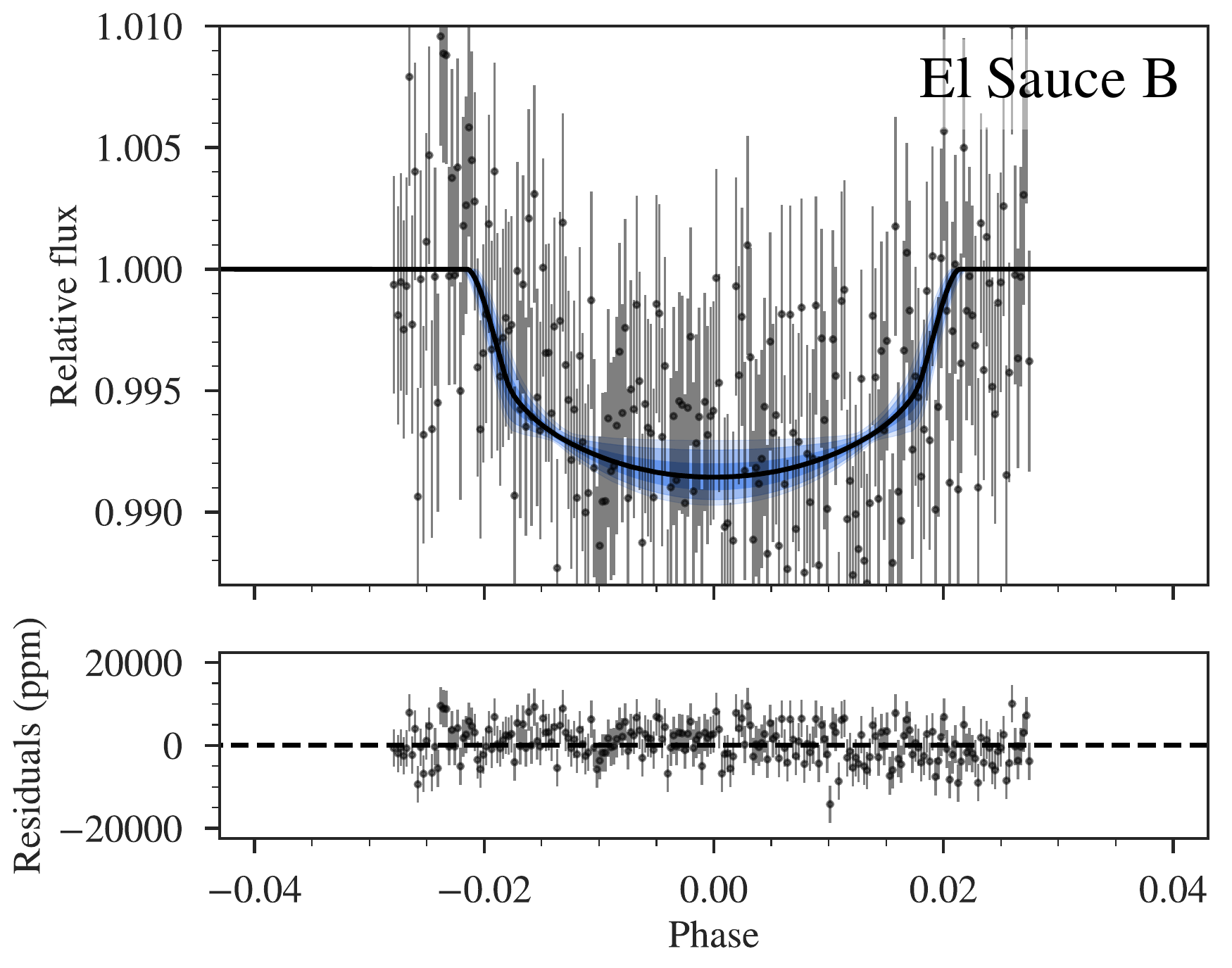}
\end{minipage}
\begin{minipage}{.33\textwidth}
  \centering
  \includegraphics[width=1\linewidth]{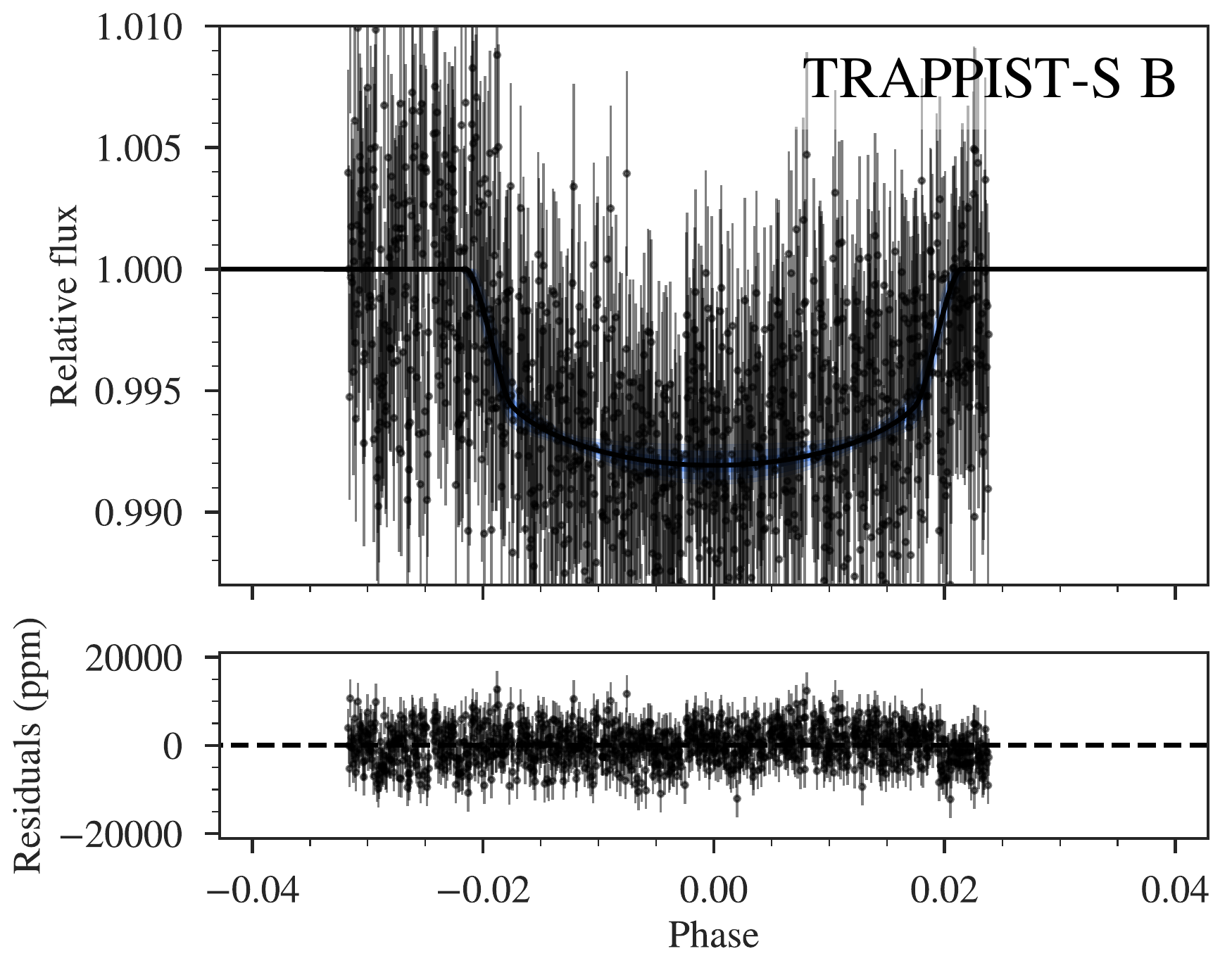}
\end{minipage}
\caption{\textit{Top}. Above is the full \textit{TESS} light curve for \tessstarnameone\ taken from Sector 1, where the best-fit model from \codename\ is overplotted (black line) along with the 68\%, 95\%, and 99\% posterior bands (blue shaded regions) taken from 5000 samples. \textit{Bottom}. Phase-folded transits for \tessstarnameone b for all available photometric instruments: \textit{TESS} (top left), LCO z band (top middle), LCO i band (top right), El Sauce (bottom left), and TRAPPIST-S (bottom right). Any GP components have been subtracted out in the phase-folded curves, and to mention specifically for the TRAPPIST-S photometry, the meridian flip had also been corrected for.}
\label{fig:toi150_phot}
\end{figure*}

\begin{figure*}
\centering
\begin{minipage}{\textwidth}
  \includegraphics[width=\linewidth]{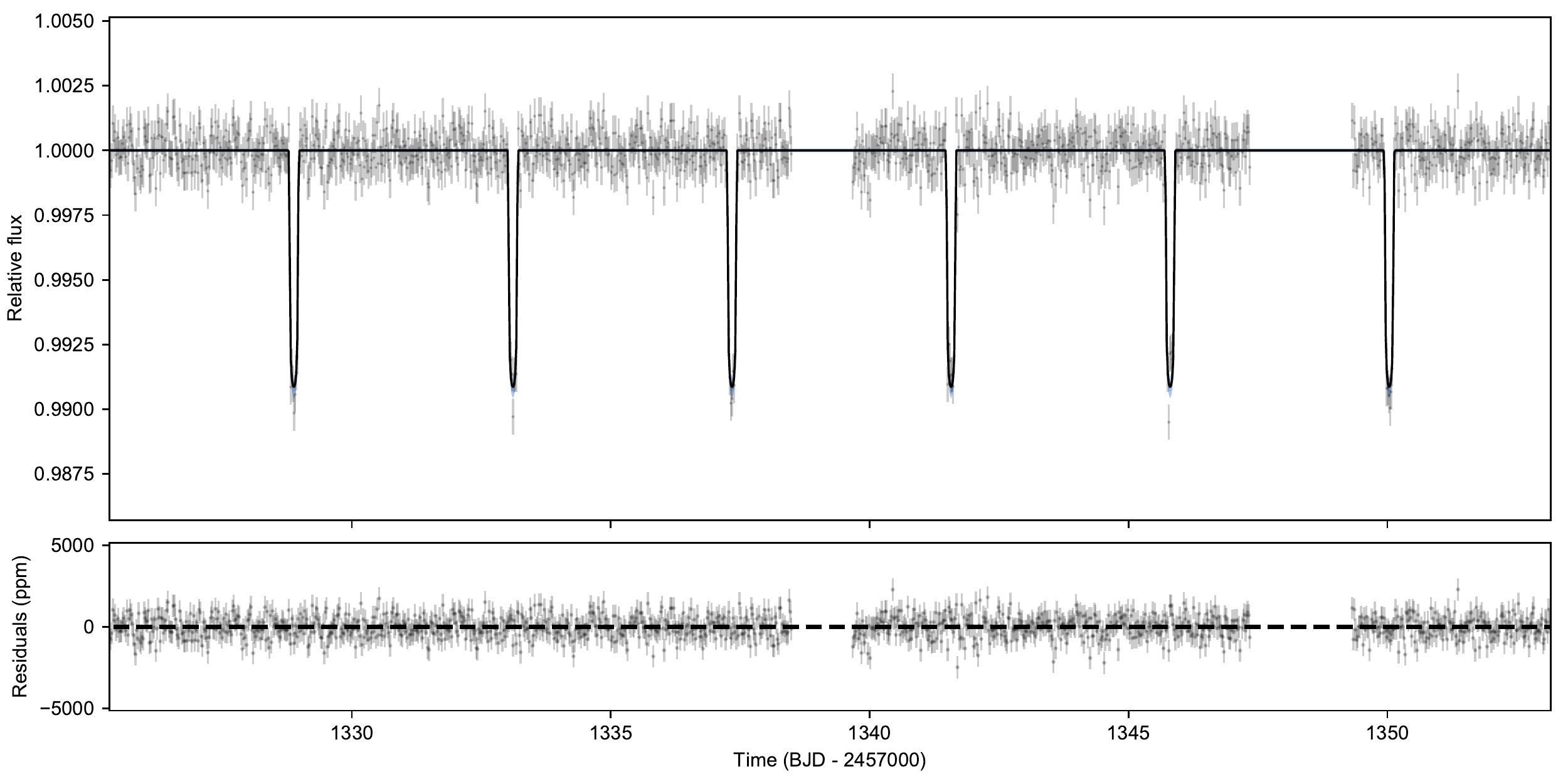}
\end{minipage}
\begin{minipage}{.33\textwidth}
  \centering
  \includegraphics[width=1\linewidth]{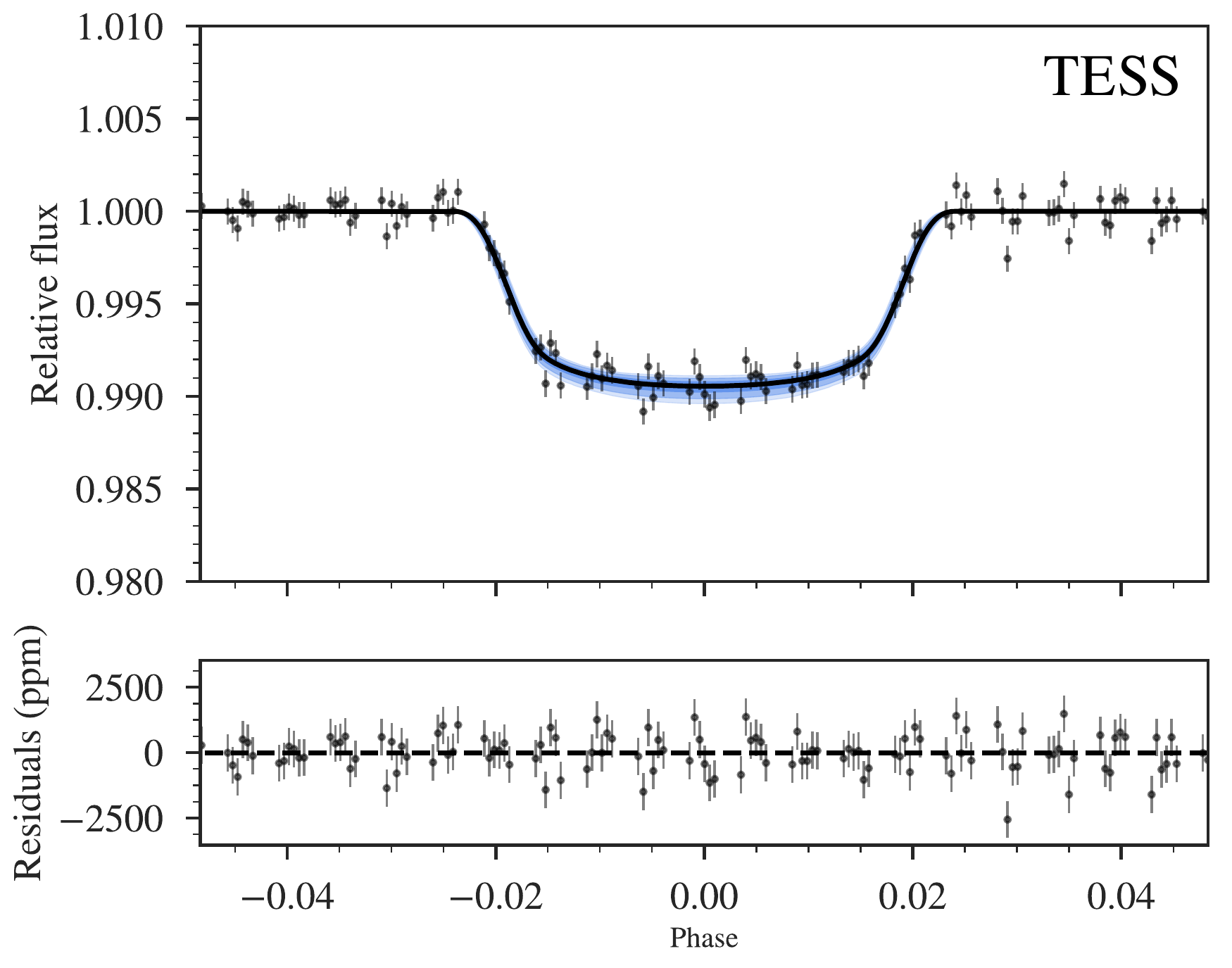}
\end{minipage}%
\begin{minipage}{.33\textwidth}
  \centering
  \includegraphics[width=1\linewidth]{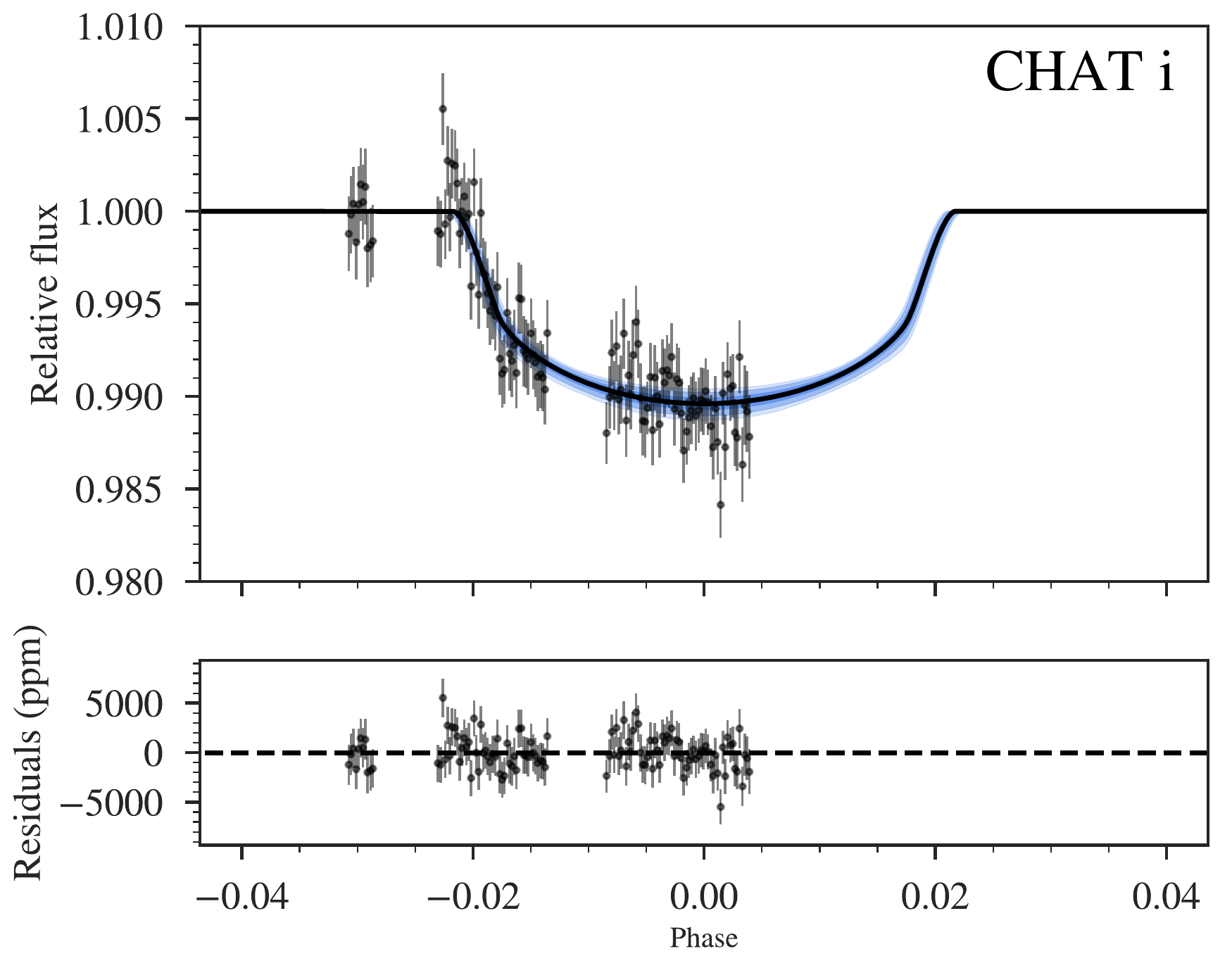}
\end{minipage}%
\begin{minipage}{.33\textwidth}
  \centering
  \includegraphics[width=1\linewidth]{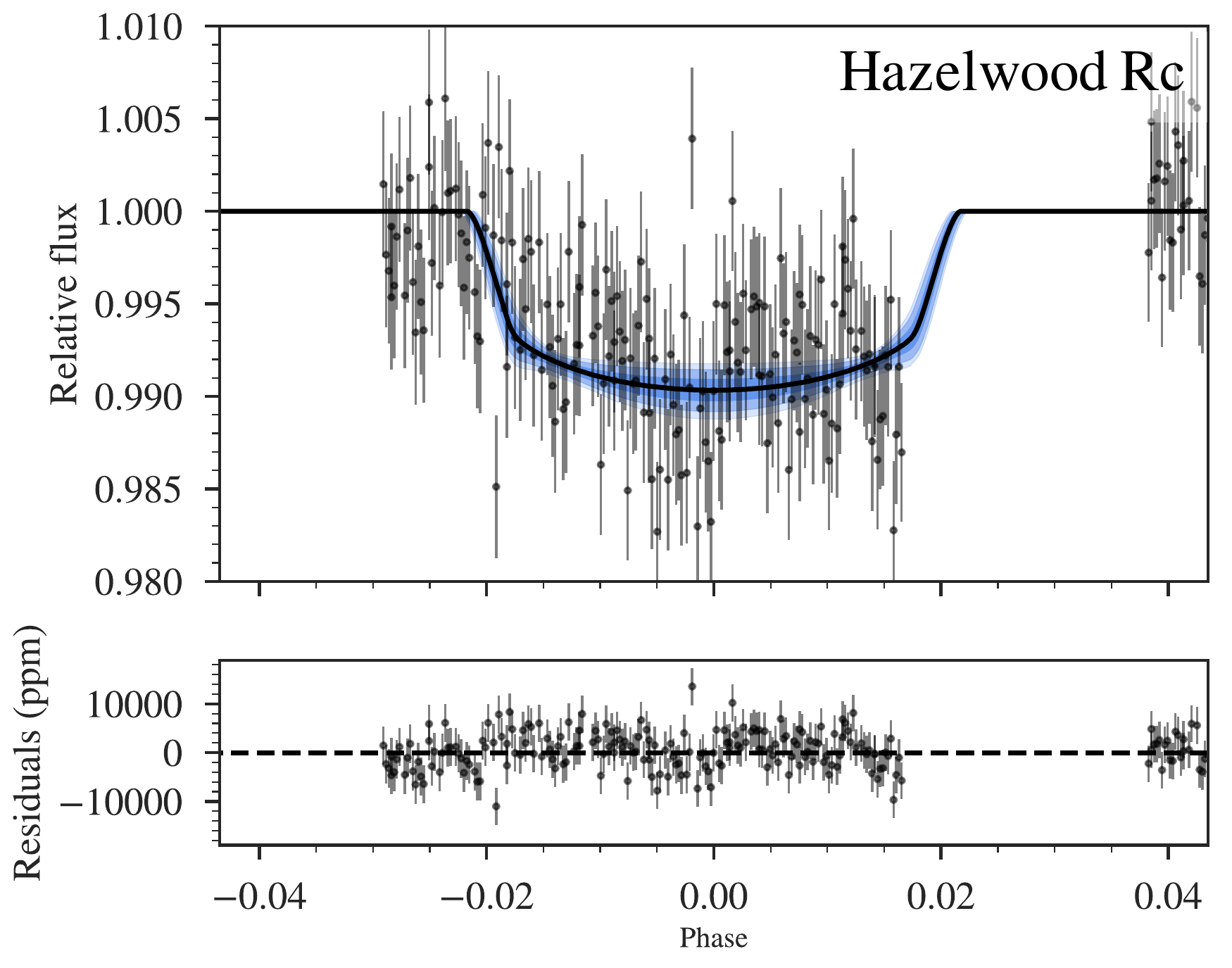}
  
\end{minipage}
\begin{minipage}{.33\textwidth}
  \centering
  \includegraphics[width=1\linewidth]{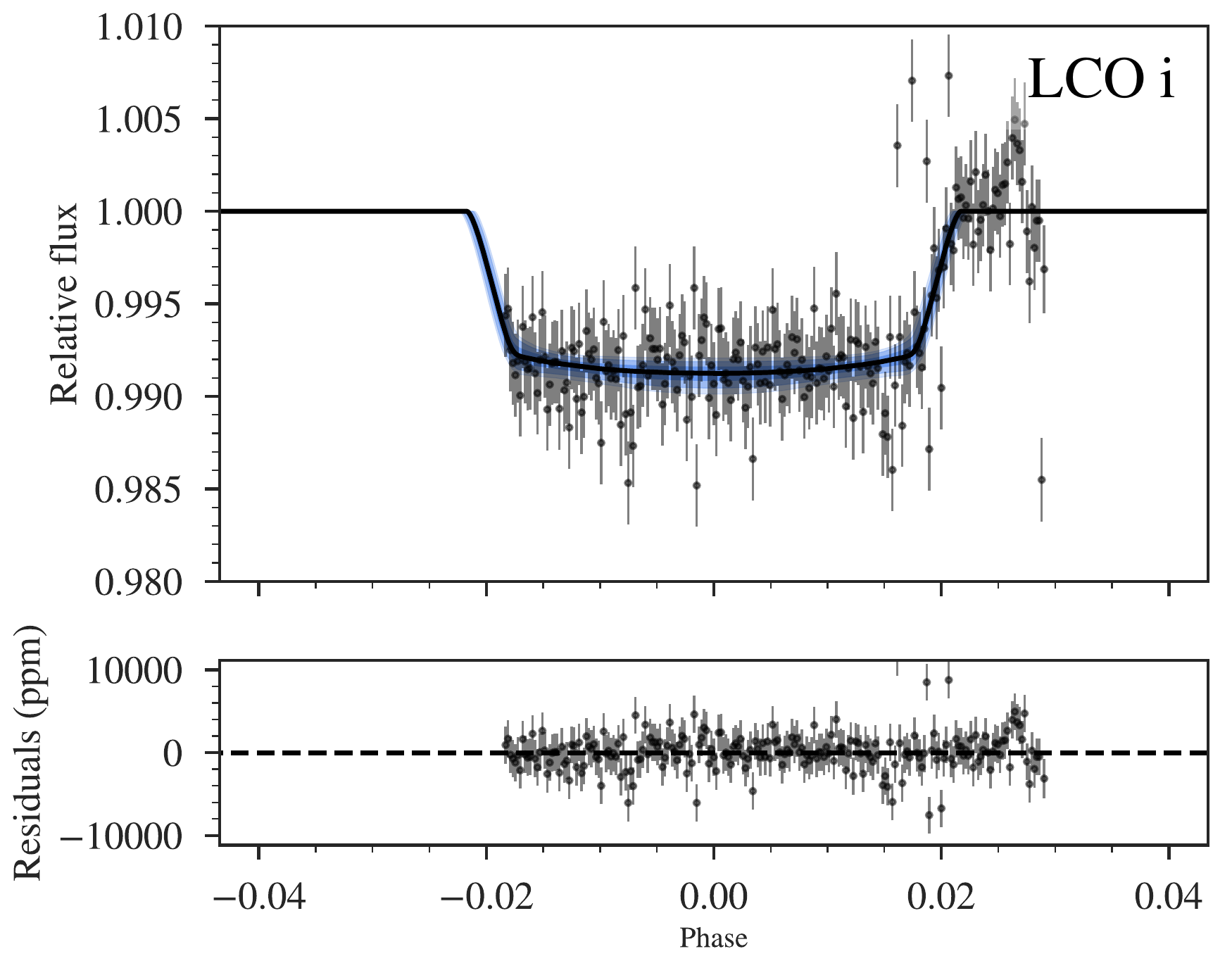}
  
\end{minipage}
\begin{minipage}{.33\textwidth}
  \centering
  \includegraphics[width=1\linewidth]{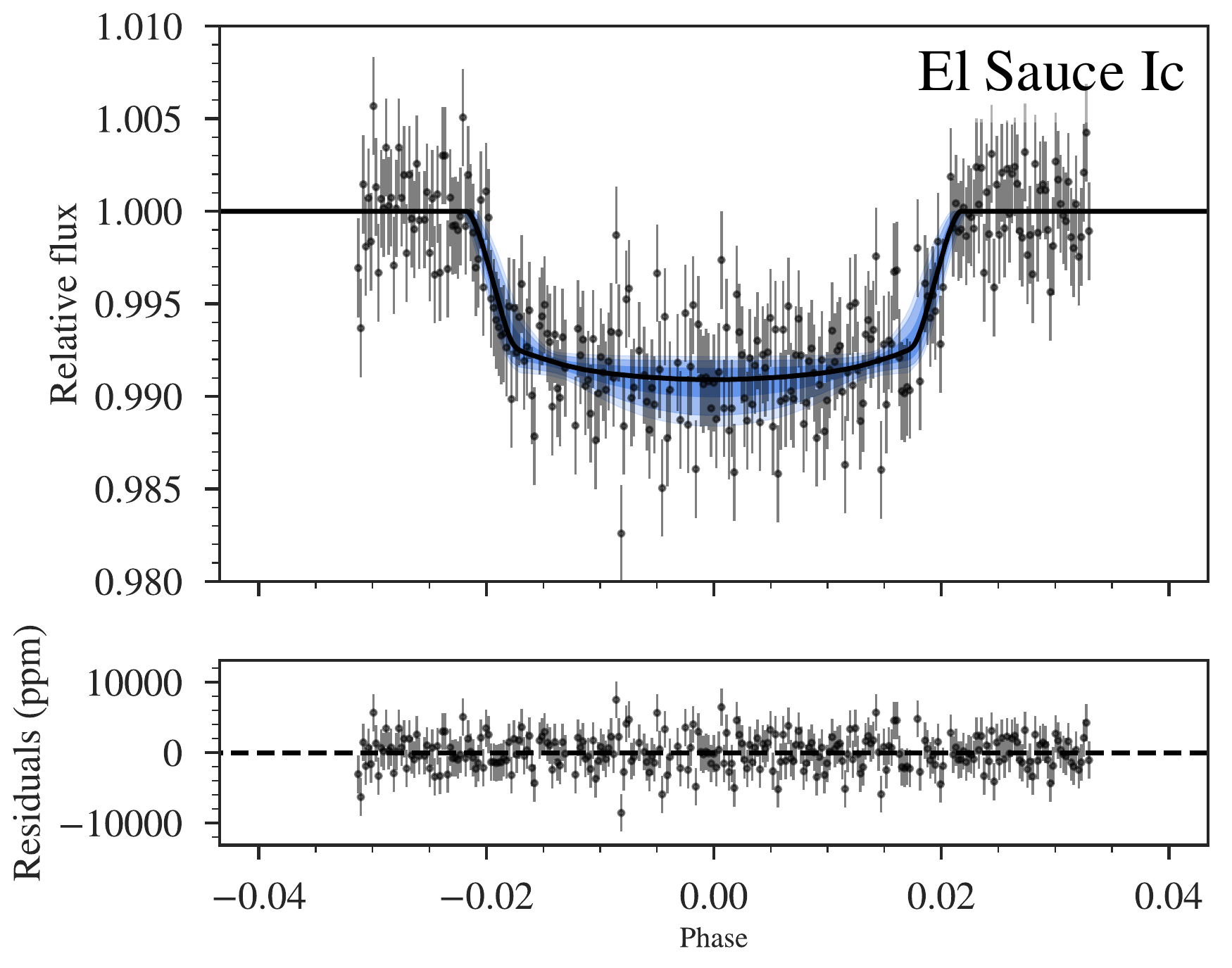}
  
\end{minipage}
\caption{\textit{Top}. Above is the full \textit{TESS} light curve for \tessstarnametwo\ taken from Sector 1, where the best-fit model from \codename\ is overplotted (black line) along with the 68\%, 95\%, and 99\% posterior bands (blue shaded regions). 5000 samples from the posterior were considered for the confidence intervals. \textit{Bottom}. Phase-folded transits for \tessstarnametwo b for all available photometric instruments: \textit{TESS} (top left), CHAT (top middle), Hazelwood (top right), LCO i band (bottom left), and El Sauce (bottom right). The phase-folded curves have been fixed by subtracting out any GP components. Gaps present in the CHAT and Hazelwood photometry can be attributed to weather and instrumental failures.}
\label{fig:toi163_phot}
\end{figure*}

Outliers that were flagged were removed as well as the same datapoints mentioned in \citep{pimen_huang} which were taken out due to the increased spacecraft pointing jitter. In order to search for possible additional signals to the ones detected by the TESS team, we analyzed the light curves using the Box-least-squares algorithm \citep[BLS;][]{kovacs:2002}. Using the whole dataset we recovered the prominent signals of \tessstarnameone b and \tessstarnametwo b of 5.87d and 4.23d, respectively. After masking these signals, no more signals are found in the photometry. In order to mitigate stellar and/or instrumental long-term trends in the photometry, we masked the in-transit data and performed a Gaussian Process (GP) regression using the quasi-periodic kernel as presented in \citep{celerite}, which we use to detrend the lightcurves of our target stars. The detrended and flattened \textit{TESS} light curves of both targets are shown in Figures \ref{fig:toi150_phot} and \ref{fig:toi163_phot}, for \tessstarnameone\ and \tessstarnametwo\ respectively, alongside with the phase-folded plots of all photometry instruments where any GP components are already subtracted. We point out that both targets exhibit photometrically quiet behavior, and therefore the pre-conditioned light curves look practically identical to the post-conditioned ones. 

Due to the large 21" pixel size of \textit{TESS}, it is imperative that ground-based follow-up phototmetry is used to confirm \textit{TESS} detections in order to avoid false positive situations, such as undiluted eclipsing binaries (i.e. the companion is not planetary but rather a low-mass star), background eclipsing binary or blended stellar binaries where the light is diluted by another star \citep{santerne:2013, desert:2015}. In addition, this is also important for studying possible transit dilutions that might give rise to wrong transit parameters if not taken into account when analyzing the \textit{TESS} photometry. We detail those follow-up photometric observations below. 

\subsection{CHAT Photometric Follow-up}
In addition to the \textit{TESS} photometry, we acquired photometric data in the i band on the night of September 21, 2018 for \tessstarnametwo\ from the Chilean-Hungarian Automated Telescope (CHAT; Jordan et al., in prep.) 0.7 meter telescope located at Las Campanas Observatory (LCO) in Chile. The primary objective of CHAT is to serve for photometric follow-up for exoplanet candidates; the telescope has achieved 1 mmag RMS precision for stars with V magnitude 12-14. The photometry was reduced with a standard pipeline which performs bias, dark and flat-field corrected images, and these were used to extract aperture photometry for various apertures. The optimal aperture was chosen as the one that, after correcting for atmospheric effects using comparison stars of similar brightness and colors, produced the light curve with the smallest root-mean-square residuals after filtering with a median filter. The resulting light curve showed an evident ingress event at the predicted time from the \textit{TESS} observations on the target. We incorporate this light curve in our joint modelling to be detailed in Section \ref{subsec:toi163analysis}.

\subsection{Hazelwood Photometric Follow-up}
Photometric follow-up data for \tessstarnametwo\ was also gathered within the \textit{TESS} Follow-up Observing Program (TFOP) Working Group; specifically, within Sub Group 1 (Seeing-limited Photometry). The data was gathered using a 0.32-m Planewave CDK telescope from Hazelwood Observatory, a backyard observatory, located in Victoria, Australia and operated by Chris Stockdale. The observed data in the Rc filter taken on October 13, 2018 included pre-transit baseline, ingress, and after-transit baseline with some missing observations around the egress. The photometry, although with large systematic trends, showed an evident ingress of the target at the expected \textit{TESS} time predicted by the \textit{TESS} observations. The aperture radius is 5.5" and there were no stars within 3' of the target with a delta magnitude less than 5.5. We incorporate this light curve as well in our joint modelling and we discuss in more detail on how to deal with the photometric variability in Section \ref{subsec:toi163_gpjitter}. One should also note that additional Hazelwood photometry for \tessstarnametwo\ was taken in the g' band on January 14, 2019, but due to cirrus cloud interference, several data points had been discarded and the quality of the remaining data would not benefit the final fit, so therefore, these datapoints were not incorporated.

\subsection{LCO/CTIO Photometric Follow-up}
Additional photometric data for \tessstarnameone\ were taken on November 9, 2018 with the 1-m telescope at Cerro Tololo Inter-American Observatory (CTIO) located near La Serena in Chile via the Las Cumbres Observatory Global Telescope (LCOGT) program \citep{brown_lco:2013}. The photometry was taken in two bands: z and i band, where both covered the egress of the transit. The aperture radius for the z band was 5.84" and showed no possible contamination from neighboring objects; whereas the aperture radius for the i band was 19.5" and showed \textit{potential} contamination. This contamination possibility was taken into consideration as a dilution factor for the fit, but it was found that the contamination is insignificant (Section \ref{subsec:toi150_fluxcontamination}).

Photometric follow-up was also taken for \tessstarnameone\ on November 12, 2018 in the i band, where the aperture radius was 13.2" and there were no apparent objects near the target with a magnitude difference less than 5.97 mag. However, there were systematics that were dealt with via 
Gaussian Process regression, as explained in Section \ref{subsec:toi163_gpjitter}.

\subsection{El Sauce Photometric Follow-up}
Data for both \tessstarnameone\ and \tessstarnametwo\ were obtained from the Observatorio El Sauce located in the R\'{i}o Hurtado Valley, in the south of the Atacama desert. \tessstarnameone\ was observed in the B filter on January 30, 2019 and \tessstarnametwo\ in the Ic filter on January 6, 2019, both covering a full transit with an aperture radius of 7.4" and using a 0.36-meter telescope. Photometry for both targets showed systematic trends that were also handled with Gaussian Process regression (see Sections \ref{subsec:toi150_gpjitter} and \ref{subsec:toi163_gpjitter}).

\subsection{TRAPPIST-South Photometric Follow-up}
Lastly, we obtained photometry for a full transit for \tessstarnameone\ on December 19, 2018 using the 0.6-meter TRAnsiting Planets and PlanetesImals Small Telescope$-$South (TRAPPIST-South) located in La Silla, Chile. Observations were carried out with good weather conditions in the B filter with an aperture radius of 5.76" and all possible candidates within 2' had been cleared. Systematics were taken care of with Gaussian Process regression, where we also accounted for a systematic jump in the flux due to a meridian flip (see Section \ref{subsec:toi150_gpjitter}).

\subsection{Gemini/DSSI Speckle Images} \label{subsec:speckle}
Speckle imaging for \tessstarnametwo\ was obtained on October 28, 2018, using the Differential Speckle Survey Instrument (DSSI) \citep{horch_dssi:2009, horch_dssi:2012, howell_dssi:2016} located at the 8-meter Gemini South Telescope at Cerro Pachon, Chile. The DSSI obtains simultaneous speckle images of targets as faint as V magnitude 16-17, in 2 channels: $R$ (692nm) and $I$ (880nm), where the spatial resolution reached is $\sim$0.017" and $\sim$0.028", respectively. The 692-nm and 880-nm filters are labeled as the $R$ and $I$ bands, respectively, since their wavelength centers align, however, the true filter is considerably narrower with a $\Delta \lambda$ of 40 nm and 50 nm for the respective wavelengths. The contrast curves (Figure \ref{fig:toi163_contrastcurve}) show that there are no stellar companions to a depth of 3.7 magnitudes for the $R$ band and 3.9 magnitudes for the $I$ band at 0.1"; and $>$4.6 and $>$5.1 magnitudes outside a radius of 0.5" for the two wavelengths, respectively. 

\subsection{FEROS Spectroscopic Follow-up}
In order to identify if the transit signals are truly due to planetary companions and to also measure the mass of the planetary companions, we obtained radial velocities (R $\approx$ 48,000) from the FEROS spectrograph \citep{feros}, which is mounted on the MPG 2.2m telescope located at La Silla Observatory in Chile. To calibrate the measurements, a simultaneous method was imposed where a ThAr calibration lamp is observed in a comparison fiber next to the science fiber, so that instrumental RV drifts can be correctly accounted for. Exposure times were on average 400-600 seconds long for these bright F-type stars. The data was reduced using the CERES pipeline \citep{ceres}.

\begin{figure*}
    \includegraphics[width=1\columnwidth]{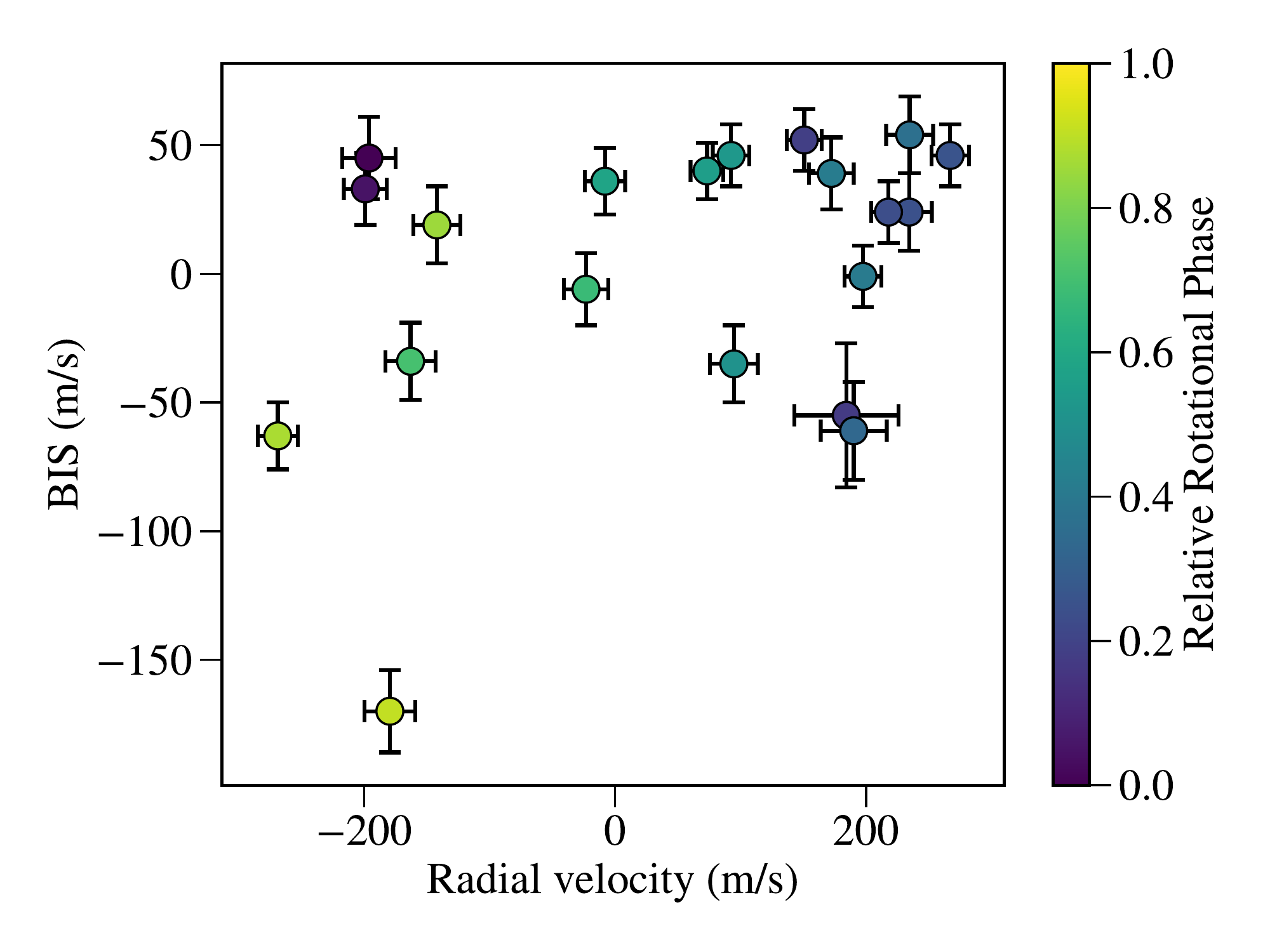}
    \includegraphics[width=1\columnwidth]{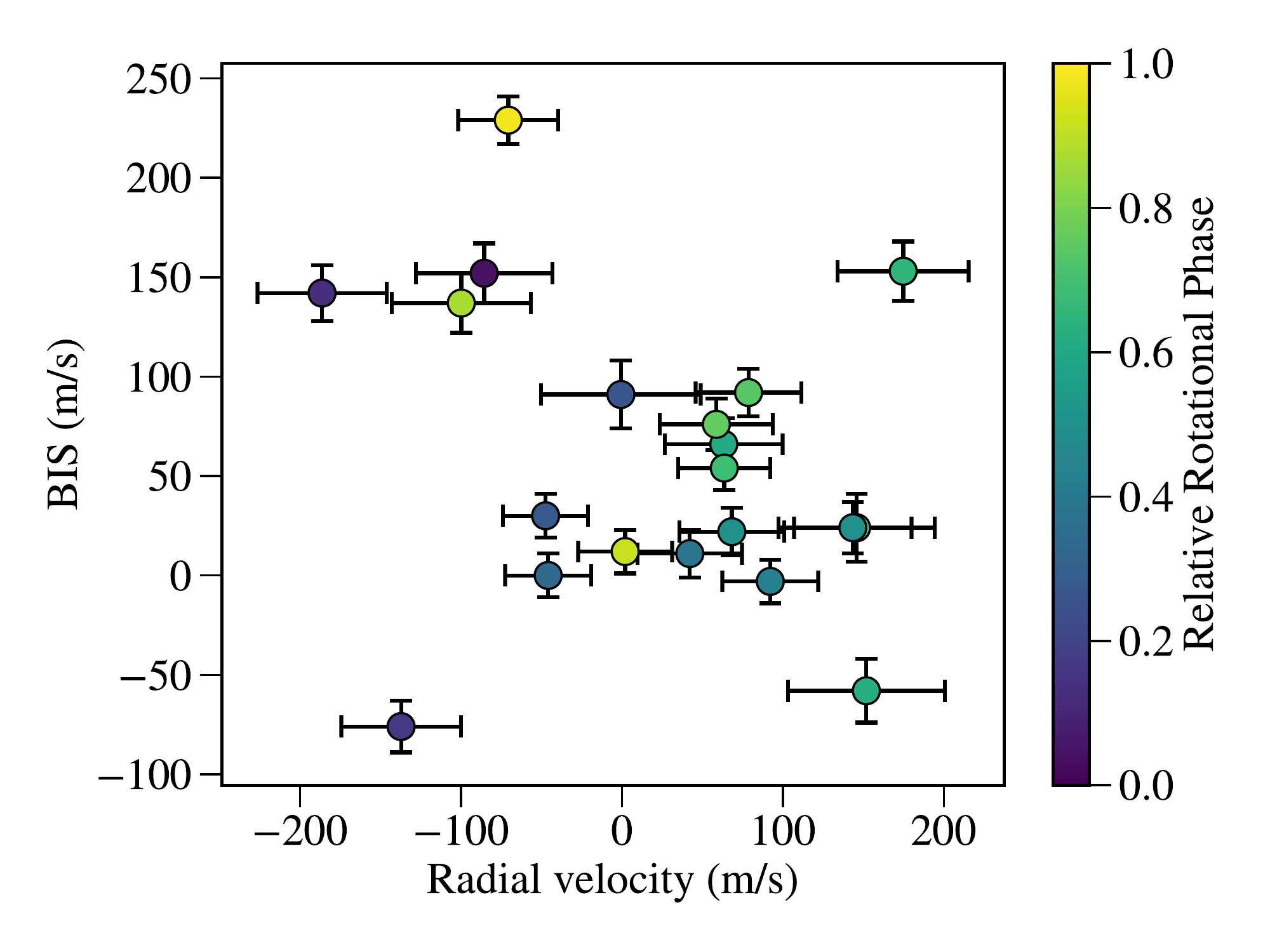}
    \caption{The radial velocity and BIS are plotted against each other for \tessstarnameone\ (left) and \tessstarnametwo\ (right) using just the FEROS data and they show no correlation. The radial velocity was offset by $\mu_{FEROS}$ (5938.91 m/s) and the color represents the phase of the period (5.8575d), both of which where taken from the posterior Tables \ref{tab:posteriors1},\ref{tab:posteriors2} for \tessstarnameone. Likewise, the radial velocity was offset by $\mu_{FEROS}$ (21393.73 m/s) and phase-folded with the period (4.231306d) given by the same posterior tables. The phase is defined to be 0 when the first data point was taken.}
    \label{fig:rvbis_correlation}
\end{figure*}

For \tessstarnameone, 20 datapoints were taken over the course of 49 days (September 19, 2018 - November 7, 2018). The data showed radial-velocities that evidently phased up with the photometric ephemerides with a semi-amplitude of 200 m/s; additionally, the stellar spectrum hinted towards a $6000$ K, $\log g = 4.0$ stellar host. Similarly, 20 datapoints were obtained for \tessstarnametwo\ over the course of 47 days (September 17, 2018 - November 3, 2018). The radial velocities also phased up with the photometric ephemerides, with a semi-amplitude of 100 m/s for the target; the stellar spectrum indicated the host star to be a $6500$ K, $\log g = 4.0$ star. No correlation was observed with the bisector spans (BIS) for any of the targets (Figure \ref{fig:rvbis_correlation}) and the data can be found in Table \ref{tab:rvdata}. 

\subsection{CORALIE Spectroscopic Follow-up}
Three high-resolution spectra were obtained for \tessstarnameone\ with CORALIE on the Swiss 1.2-m Euler telescope at La Silla Observatory, Chile \citep{Queloz2000} over a timespan of 32 days (October 6, 2018 - November 7, 2018). CORALIE has resolution $R =$ 60,000 and uses simultaneous Fabry-P\'{e}rot wavelength calibration during science exposures. The science-fibre is 2\arcsec\ on sky. For each epoch we compute the RVs by cross-correlation with a binary G2 mask using the standard CORALIE pipeline. Line-profile diagnostics such as bisector span and FWHM are produced as well, to check for correlations with RV of which none were found. We also compute RVs using other binary masks ranging from A0 to M4, to check for a mask-dependent signal indication a blend. The CORALIE RVs confirm the planetary nature of the \textit{TESS} detetction and is in phase with the transit ephemerides.

\begin{figure*}
\centering
\begin{minipage}{.55\textwidth}
  \centering
  \includegraphics[width=1\linewidth]{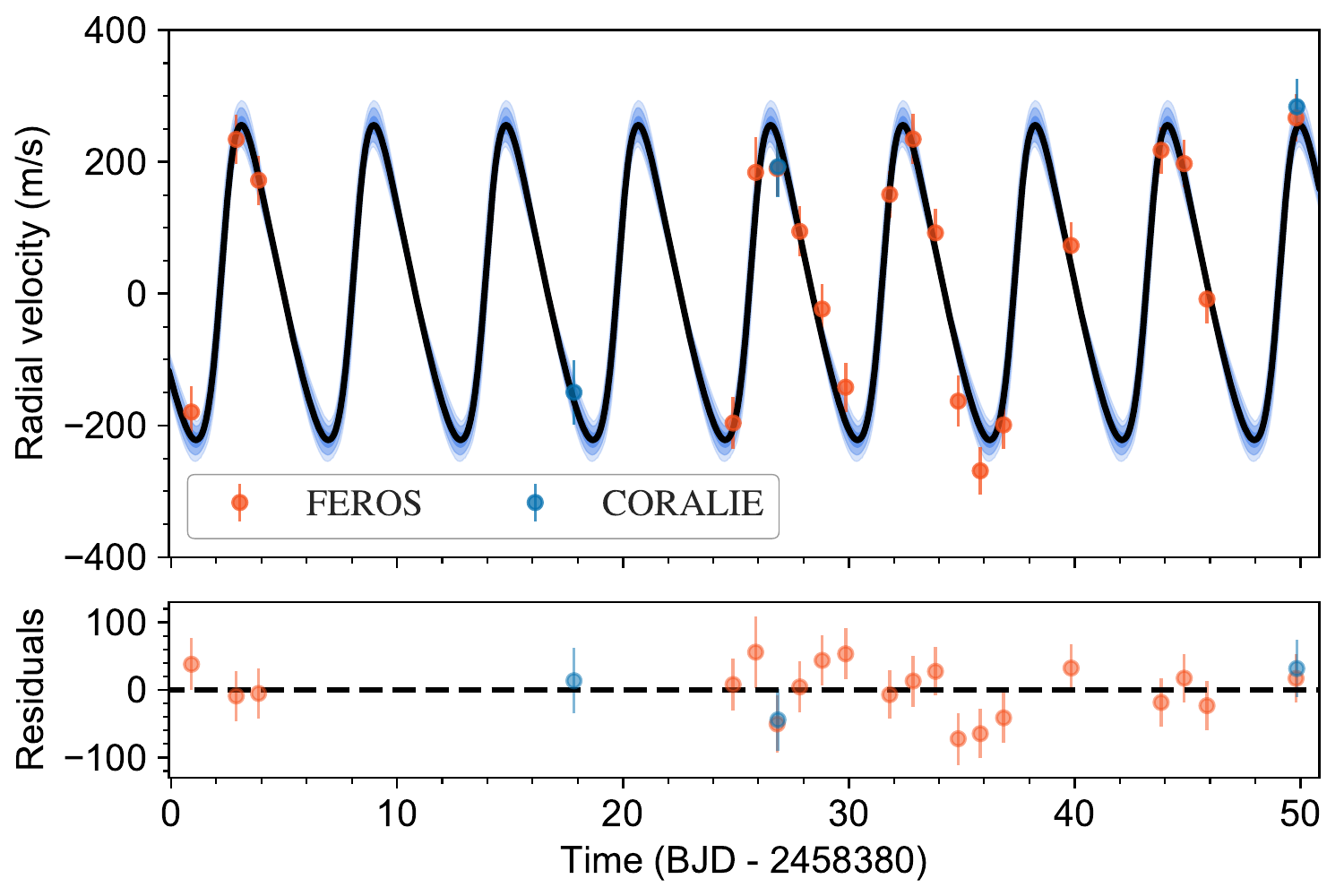}
\end{minipage}
\begin{minipage}{.4\textwidth}
  \centering
  \includegraphics[width=.95\linewidth]{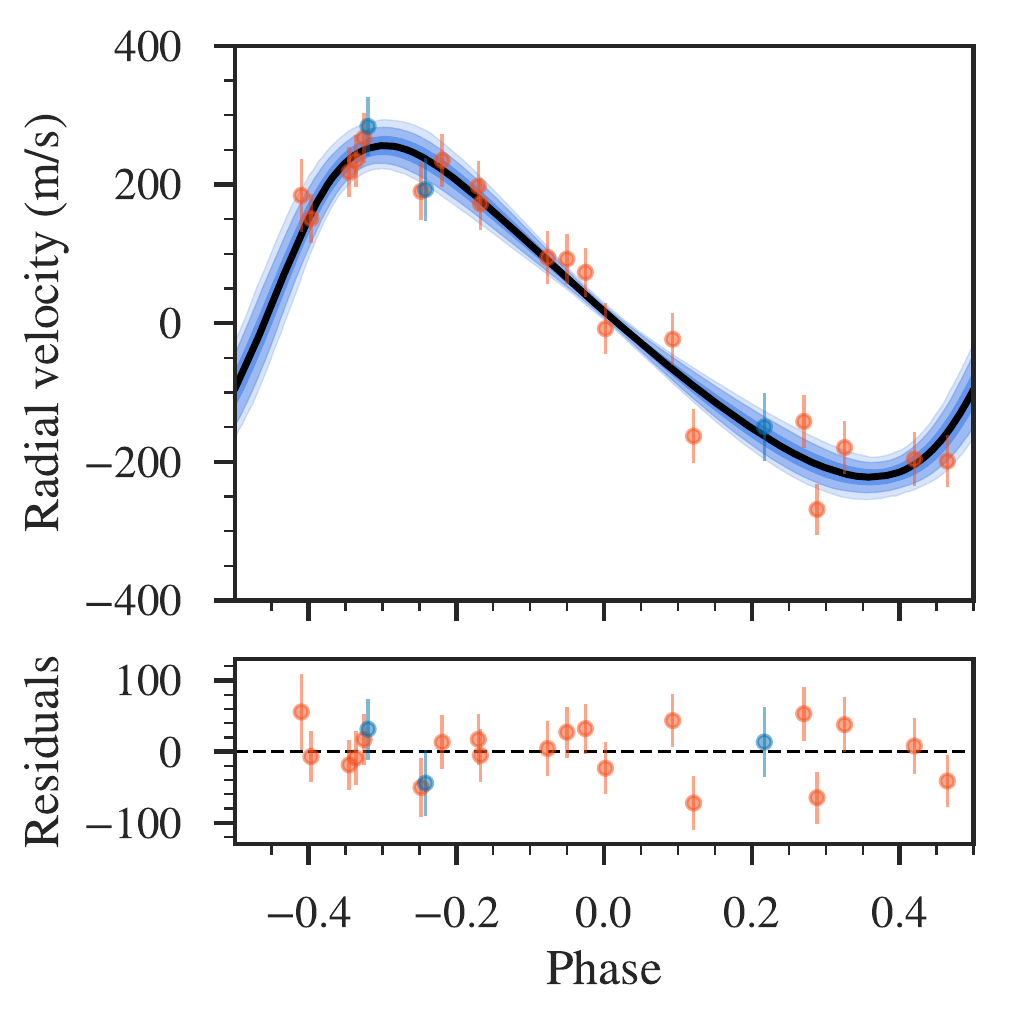}
\end{minipage}
\caption{\textit{Left}. The radial velocity measurements for \tessstarnameone\ are illustrated, along with the best model (black line) and the 68\%, 95\%, and 99\% posterior bands (blue bands) using 5000 samples from the posteriors. FEROS and CORALIE data points are shown in orange and blue, respectively. Below are then the residuals after subtracting the best model fit. \textit{Right}. The phased radial velocity measurements for \tessstarnameone b, where one can see the eccentric behavior of the signal's orbit.}
\label{fig:toi150_rvs}
\end{figure*}

\begin{figure*}
\centering
\begin{minipage}{.55\textwidth}
  \centering
  \includegraphics[width=1\linewidth]{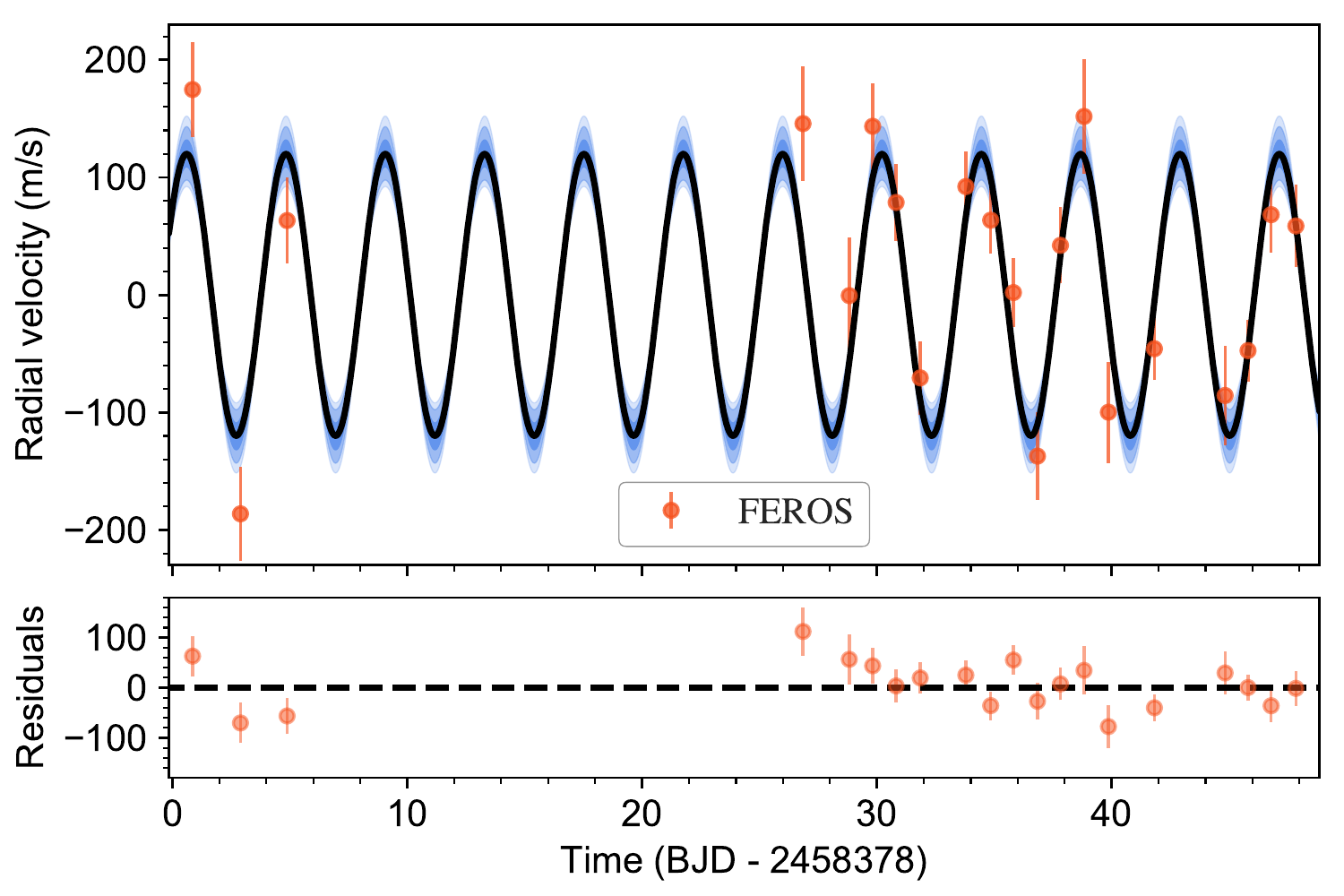}
\end{minipage}
\begin{minipage}{.4\textwidth}
  \centering
  \includegraphics[width=.95\linewidth]{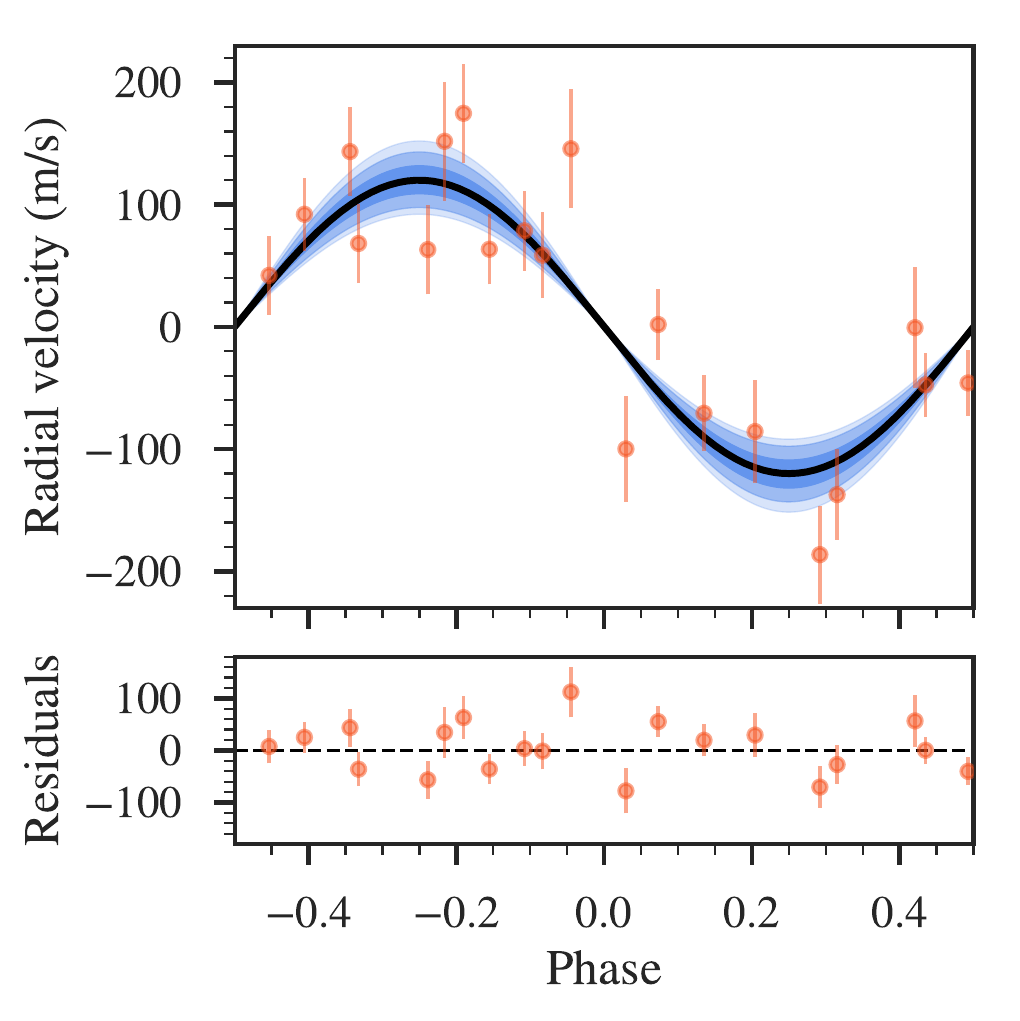}
\end{minipage}%
\caption{\textit{Left}. The FEROS radial velocity measurements for \tessstarnametwo\ are presented, along with the best model (black line) and the 68\%, 95\%, and 99\% posterior bands (blue bands) based on 5000 samples.  Below are then the residuals after subtracting the best model fit. \textit{Right}. The phased radial velocity measurements for \tessstarnametwo b.}
\label{fig:toi163_rvs}
\end{figure*}

\begin{figure*}
	\includegraphics[height=0.8\columnwidth]{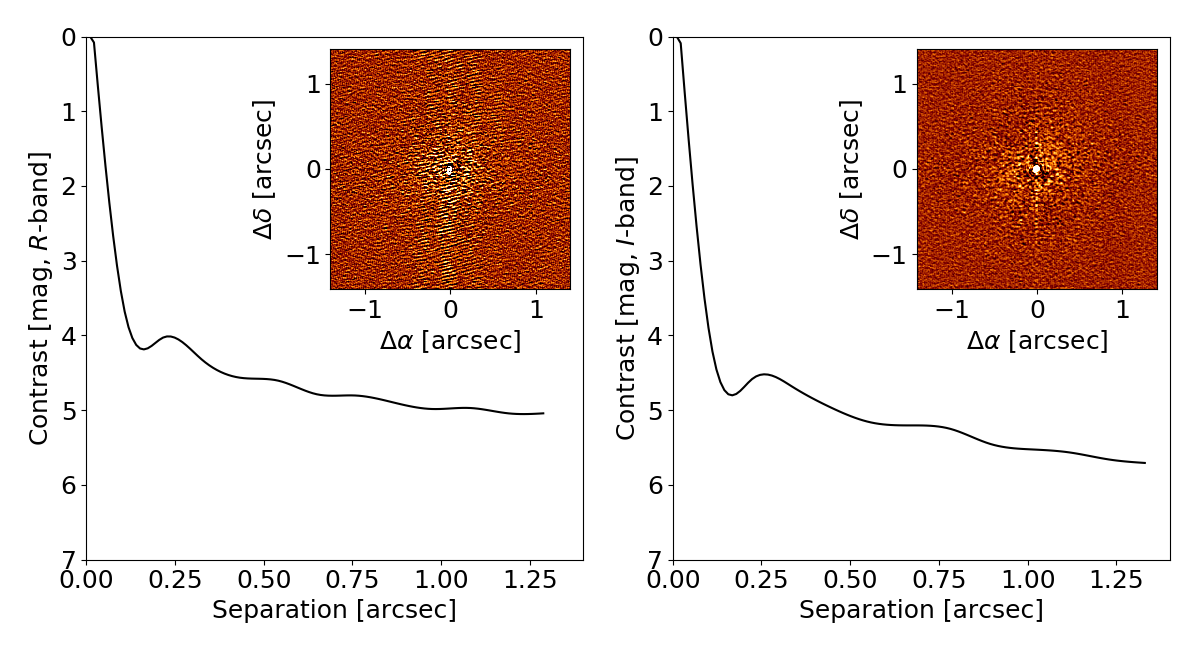}
	\caption{Presented here are the Gemini speckle interferometric observation contrast curves for the $R$ (692nm) and $I$ (880nm) band for \tessstarnametwo, along with the reconstructed images to show that there are no close stellar companions that could affect the light curve.}
	\label{fig:toi163_contrastcurve}
\end{figure*}

\section{Analysis} \label{sec:analysis}
\subsection{Stellar Parameters} \label{sec:stellarparams}
To derive the stellar parameters for the host stars of these two targets, we analyzed the co-added FEROS spectra via the Zonal Atmospheric Stellar Parameters Estimator algorithm \citep[ZASPE,][]{brahm:2015,brahm:2016:zaspe}. This code computes the atmospheric parameters ($T_{eff}$, log$g$, [Fe/H]) and the projected rotational velocity ($v\sin i$) by comparing the observed spectra to a grid of synthetic models generated from the ATLAS9 model atmospheres \citep{atlas9}. Only spectral regions that are significantly sensitive to changes in the atmospheric parameters are used for comparison. This process is then executed in an iterative method, where the uncertainties are obtained through Monte Carlo simulations. With this procedure we find that \tessstarnameone\ has an effective temperature of $T_{\textnormal{eff}}=6255.0 \pm 90.0$ K, a surface gravity of $\log g=4.20\pm0.0090$ dex, a metallicity of $[\textnormal{Fe}/\textnormal{H}]=0.28 \pm 0.036$ dex, and a projected rotational velocity of $v\sin i=7.96\pm0.28$ m s$^{-1}$. As for \tessstarnametwo, we find an effective temperature of $T_{\textnormal{eff}}=6495.0\pm90.0$ K, a surface gravity of $\log g=4.187\pm0.011$ dex, a metallicity of $[\textnormal{Fe}/\textnormal{H}]=0.220\pm0.041$ dex, and a projected rotational velocity of $v\sin i=14.08\pm0.27$ m s$^{-1}$. 

We then followed the two step procedure adopted in \cite{k2-232, k2-161} to infer the physical parameters and evolutionary stage of the host stars. First, we are able to derive a very precise stellar radius of each star by combining the parallax measurement provided by \textit{Gaia} DR2 with public broad band photometric measurements (taken from Tycho-2 or 2MASS).
Then we use the Yonsei-Yale isochrones  \citep{yi:2001} to estimate the stellar mass and age of each host star by comparing the obtained effective temperature and stellar radius to those predicted by the isochrones. In the end, we obtain radius values of $R = 0.012$ for \tessstarnameone\ and $R = 1.648^{+0.023}_{-0.025}$ for \tessstarnametwo; and then mass values of $M=1.351^{+0.038}_{-0.026}$ and $M=1.4352^{+0.029}_{-0.028}$ for the stars, respectively. From there, we can compute the stellar density, $\rho_*$, for which we will be using as a prior for the fits.
The derived stellar parameters can be found in Table \ref{tab:stellarparams}.

\begin{table*}
 \centering 
 \begin{center}
\caption{Stellar parameters of \tessstarnameone\ and \tessstarnametwo.}
 \label{tab:stellarparams}
 \begin{threeparttable}
  \centering
  \begin{tabular}{ lccr }
   \hline
   \hline
     Parameter &  \tessstarnameone\ Value &  \tessstarnametwo\ Value & Source\\
   \hline
Identifying Information\\
~~~TIC ID & 271893367 &  179317684 & TIC$^a$\\
~~~GAIA ID & 5262709709389254528 & 4651366259202463104 &\textit{Gaia} DR2$^b$ \\
~~~2MASS ID & J07315176-7336220 & J05190435-7153441 & 2MASS$^c$\\
~~~R.A. (J2015.5, h:m:s) & $7^h31^m51.7^s$ & $5^h19^m4.3^s$ & \textit{Gaia} DR2$^b$ \\ 
~~~DEC (J2015.5, d:m:s) & $-73^\circ36'21.73''$ & $-71^\circ53'43.9''$ & \textit{Gaia} DR2$^b$\\ 
Proper motion and parallax \\
~~~$\mu_\alpha \cos \delta$ (mas yr$^{-1}$) & 27.14 $\pm$ 0.03&  7.14 $\pm$ 0.07 & \textit{Gaia} DR2$^b$ \\
~~~$\mu_\delta$ (mas yr$^{-1}$) & -15.21 $\pm$ 0.03 & 16.37 $\pm$ 0.08 & \textit{Gaia} DR2$^b$ \\
~~~Parallax (mas) & 2.94 $\pm$ 0.02 & 2.40 $\pm$ 0.05 & \textit{Gaia} DR2$^b$ \\
Spectroscopic properties\\
~~~$T_\textnormal{eff}$ (K) & $6255 \pm 90$& $6495 \pm 90$ & ZASPE$^d$\\
~~~Spectral Type & F & F & ZASPE$^d$\\
~~~[Fe/H] (dex) & $0.28\pm 0.036$ & $0.22\pm 0.041$ & ZASPE$^d$\\
~~~$\log g_*$ (cgs)& $4.13\pm 0.009$ & $4.187\pm 0.011$ & ZASPE$^d$\\
~~~$v\sin(i_*)$ (km/s)& $7.96\pm 0.279$ & $14.08\pm 0.266$ & ZASPE$^d$\\
Photometric properties\\\
~~~$T$ (mag)& $10.865 \pm 0.019$ & $10.843 \pm 0.018$ & TIC$^a$\\
~~~$G$ (mag)& $11.34 \pm 0.015$ & $11.22 \pm 0.015$ & \textit{Gaia} DR2$^b$ \\
~~~$B$ (mag)& $12.173\pm 0.212$ & $11.852\pm 0.204$ & Tycho-2$^e$\\
~~~$V$ (mag)& $11.39\pm0.0015$ & $11.467 \pm0.0014$ &Tycho-2$^e$\\
~~~$J$ (mag)& $10.324\pm0.028$ & $10.404\pm0.021$ & 2MASS$^c$\\
~~~$H$ (mag)& $10.045\pm0.022$ & $10.153\pm0.024$ & 2MASS$^c$\\
~~~$Ks$ (mag)& $9.94\pm0.019$ & $10.124\pm0.023$ & 2MASS$^c$\\
Derived properties\\
\vspace{0.1cm}
~~~$M_*$ ($M_\odot$)& $1.351^{+0.038}_{-0.026}$ & $1.4352^{+0.029}_{-0.028}$ & YY$^{*}$\\
~~~$R_*$ ($R_\odot$)& $1.526^{+0.012}_{-0.012}$ & $1.648^{+0.023}_{-0.025}$ & YY$^{*}$\\
~~~$L_*$ ($L_\odot$)& $3.137^{+0.340}_{-0.270}$ & $4.330^{+0.250}_{-0.256}$ & YY$^{*}$\\
~~~$M_V$ & $3.507^{+0.107}_{-0.153}$ & $3.125^{+0.069}_{-0.072}$ & YY$^{*}$\\
~~~Age (Gyr)& $2.346^{+0.425}_{-0.901}$ & $1.823^{+0.300}_{-0.331}$ & YY$^{*}$\\
~~~$\rho_*$ (kg m$^{-3}$)& $533.2^{+14.4}_{-16.5}$ & $451.8^{+18.9}_{-19.4}$ & YY$^{*}$\\
   \hline
   \end{tabular}
      \textit{Note}. Logarithms given in base 10.\\
    (a) \textit{TESS} Input Catalog \citep{stassun_tesslist:2018}; (b) \textit{Gaia} Data Release 2 \citep{gaiadr2summary}; (c) Two-micron All Sky Survey \citep{2mass}; (d) Zonal Atmospheric Stellar Parameters Estimator \citep{brahm:2015,brahm:2016:zaspe}; (e) Tycho-2 Catalog \citep{tycho}\\
      *: Yonsei-Yale isochrones \citep{yi:2001}; using stellar parameters obtained from ZASPE.
 \end{threeparttable}
 \end{center}
 \end{table*}
 
\subsection{Joint Analysis} \label{subsec:joint}

For both \textit{TESS} targets, a simultaneous analysis of the photometry, radial velocity, and stellar density was efficiently preformed using a new algorithm, \codename\ \citep{juliet:2018}, as applied in two other \textit{TESS} discovery papers \citep{brahms_newpaper, espinoza_tess:2019}. \codename\ makes use of makes use of Nested Samplers using either MultiNest \citep{multinest} via the \texttt{PyMultiNest} package \citep{pymultinest} or the \texttt{dynesty} package \citep{dynesty:2018} in order to allow 
the computation of Bayesian model log-evidences, $\ln Z$, useful for model comparison. This new algorithm also employs \texttt{batman} \citep{batman} for modeling the transit data and \texttt{radvel} \citep{radvel} for modeling the radial velocities. This includes the ability to fit multiple transiting and non-transiting planets, combining a variety of photometric and radial velocity data sets where each would have its own Gaussian Process hyperparameters or commonly shared hyperparameters, if desired.

The advantage of this joint-modeling code, \codename, is its versatility where we can fit a variety of parameters efficiently and explore the parameter space fully given that we are implementing a nested sampling algorithm. Instead of starting off with an initial parameter vector around a likelihood maximum found via optimization techniques, as done in common sampling methods, nested sampling samples straight from the given priors. This would mean that large priors would take computationally more time; for this reason, our prior choices have been selected to be the ideal balance between being informed, 
yet wide enough to fully acquire the posterior distribution map. 

As mentioned above, \codename\ lets us perform model comparison (e.g., eccentric versus circular orbits, or N-planet models versus N+1-planet models) by comparing the differences in Bayesian log-evidences, $\Delta \ln Z$. We follow the rule-of-thumb here that if $\Delta \ln Z$ $\lesssim$ 3, then the models are indistinguishable and neither is preferred so the simpler model would then be chosen. For any $\Delta \ln Z$ that is greater than 3, the model with the larger Bayesian log-evidence is favored. 

The specific details of the analysis for each target, \tessstarnameone\ and \tessstarnametwo, are outlined in Sections \ref{subsec:toi150analysis} and \ref{subsec:toi163analysis} respectively. In general, however, the same steps were more or less taken with some minor differences regarding eccentricities, instrument jitter terms ($\sigma_w$), and instrument dilution factors ($D$). The treatment of the \textit{TESS} lightcurves for both targets was identical in the sense that they are long-cadence observations, so therefore, we applied a resampling technique \citep[outlined in][]{binningissinning}, where we resampled $N=20$ points per given datapoint. In order to avoid potential biases by our limb-darkening assumptions \citep[see, e.g.,][]{espinoza_jordan_limb:2015}, we choose to fit for the limb-darkening coefficients simultanously on our transit fitting procedure. The \textit{TESS} photometry was modelled with a quadratic limb-darkening law, whereas the other photometric instruments were assigned linear limb-darkening laws \citep[both parametrized with the uniform sampling scheme of][]{kipping_limb:2013}. The selection of a two-parameter law for precise space-based instruments like TESS and of the linear law for the ground-based instruments was based on the work of \cite{espinoza_jordan_limb:2016}. Furthermore, instead of fitting directly for the planet-to-star radius ratio ($p=R_p/R_*$) and the impact parameter of the orbit ($b$), we choose to use the parametrization introduced in \cite{Espinozapb:2018} in which we fit for the parameters  $r_1$ and $r_2$ which ensure we explore the whole range of physically plausible values in the ($p,b$) plane. 
Our final fits include $\rho_*$, the stellar density as taken from Table \ref{tab:stellarparams}, as a prior, largely due to newer and more precise data (i.e. from \textit{Gaia} Data Release 2), since we can now take advantage of the estimated stellar density and use it to constrain $P$ and $a/R_*$ of single transiting planets, instead of the opposite as previously done \citep{Seager:2003,sozzetti:2007}. The impact that a stellar density prior may have on various parameters is discussed in Section \ref{sec:stellardensitycomp}.

Before creating the final joint fits for each target, individual fits on photometry-only and radial velocity-only were first carried out. The posteriors from these fits were then taken into consideration when setting up the priors for the joint fit. The main differences in priors between the joint fits and the individual fits, were that the $P$ and $t_0$ parameters were adapted to have a normal prior rather than a uniform prior where the normal prior is based on the posterior distributions from the transit-only fit. The prior for the semi-amplitude $K$, was kept to be uniform but more constrained than searching through the entire parameter space. 

\begin{table*}
    \centering
    \caption{Below are the priors used for \tessstarnameone\ and \tessstarnametwo\ for the \textit{final} joint analysis fit using \codename.
    As a recollection, $p=R_p/R_*$ and $b=(a/R_*)\cos(i_p)$, where $R_p$ is the planetary radius, $R_*$ the stellar radius, $a$ the semi-major axis of the orbit and $i_p$ the inclination of the planetary orbit with respect to the plane of the sky. $e$ and $\omega$ are the eccentricity and argument of periastron of the orbits. The prior labels of $\mathcal{N}$, $\mathcal{U}$, and $\mathcal{J}$ represent normal, uniform, and Jeffrey's distributions. Reasons for why some parameters are fixed to a value are better explained in detail in Sections \ref{subsec:toi150_gpjitter} and \ref{subsec:toi163_gpjitter}. The parametrization for $(p,b)$ using $(r_1,r_2)$ \citep{espinoza_jordan_limb:2015, espinoza_jordan_limb:2016} and the linear $(q_1)$ and quadratic $(q_1,q_2)$ limb-darkening parametrization \citep{kipping_limb:2013} are both described in Section \ref{subsec:joint}. Continuation of the priors are in Table \ref{tab:priors2}.}  
    \label{tab:priors1}
    \begin{tabular}{lcccl} 
        \hline
        \hline
        Parameter name & Prior (\tessstarnameone b) & Prior (\tessstarnametwo b) & Units & Description \\
                \hline
                \hline        
        Parameters for the star & & \\
        ~~~$\rho_*$ & $\mathcal{N}(535.76,17.48^2)$ & $\mathcal{N}(451.406,29.5^2)$ & kg/m$^3$ & Stellar density. \\
        \hline
        Parameters for planet b\\
        ~~~$P_b$ & $\mathcal{N}(5.87,0.01^2)$ & $\mathcal{N}(4.23,0.001^2)$ & days & Period. \\
        ~~~$t_{0,b} - 2458320$ & $\mathcal{N}(6.32,0.01^2)$ & $\mathcal{N}(8.88,0.01^2)$ & days & Time of transit-center. \\
        ~~~$r_{1,b}$ & $\mathcal{U}(0,1)$ & $\mathcal{U}(0,1)$ & --- & Parametrization for $p$ and $b$. \\
        ~~~$r_{2,b}$ & $\mathcal{U}(0,1)$ & $\mathcal{U}(0,1)$ & --- & Parametrization for $p$ and $b$. \\
        ~~~$K_{b}$ & $\mathcal{U}(150,300)$ & $\mathcal{U}(80,170)$ & m/s & Radial-velocity semi-amplitude. \\
        ~~~$\mathcal{S}_{1,b} = \sqrt{e_b}\sin \omega_b$ & $\mathcal{U}(-1,1)$ & 0.0 (fixed) & --- & Parametrization for $e$ and $\omega$. \\
        ~~~$\mathcal{S}_{2,b} = \sqrt{e_b}\cos \omega_b$ & $\mathcal{U}(-1,1)$ & 0.0 (fixed) & --- & Parametrization for $e$ and $\omega$. \\
        
        \hline
        Parameters for TESS\\
        ~~~$D_{\textnormal{TESS}}$ & $\mathcal{U}(0.5,1)$ & $\mathcal{U}(0,1)$ & --- & Dilution factor for TESS. \\
        ~~~$M_{\textnormal{TESS}}$ & $\mathcal{N}(0,0.1^2)$ & $\mathcal{N}(0,0.1^2)$ & ppm & Relative flux offset for TESS. \\
        ~~~$\sigma_{w,\textnormal{TESS}}$ & $\mathcal{J}(0.1,600^2)$ & 0.0 (fixed) & ppm & Extra jitter term for \textit{TESS} lightcurve. \\
        ~~~$q_{1,\textnormal{TESS}}$ & $\mathcal{U}(0,1)$ & $\mathcal{U}(0,1)$ & --- & Quadratic limb-darkening parametrization. \\
        ~~~$q_{2,\textnormal{TESS}}$ & $\mathcal{U}(0,1)$ & $\mathcal{U}(0,1)$ & --- & Quadratic limb-darkening parametrization. \\
        
        \hline
        Parameters for CHAT \\
        ~~~$D_{\textnormal{CHAT}}$ & --- & 1.0 (fixed) & --- & Dilution factor for CHAT. \\
        ~~~$M_{\textnormal{CHAT}}$ & --- & $\mathcal{N}(0,0.1^2)$ & ppm & Relative flux offset for CHAT. \\
        ~~~$\sigma_{w,\textnormal{CHAT}}$ & --- & 0.0 (fixed) & ppm & Extra jitter term for CHAT lightcurve. \\
        ~~~$q_{1,\textnormal{CHAT}}$ & --- & $\mathcal{U}(0,1)$ & --- & Linear limb-darkening parametrization. \\
        \hline
        Parameters for Hazelwood \\
        ~~~$D_{\textnormal{Hazwelwood}}$ & --- & 1.0 (fixed) & --- & Dilution factor for Hazwelwood. \\
        ~~~$M_{\textnormal{Hazwelwood}}$ & --- & $\mathcal{N}(0,0.1^2)$ & ppm & Relative flux offset for Hazwelwood. \\
        ~~~$\sigma_{w,\textnormal{Hazwelwood}}$ & --- & $\mathcal{J}(0.1,5000^2)$ & ppm & Extra jitter term for Hazwelwood lightcurve. \\
        ~~~$q_{1,\textnormal{Hazwelwood}}$ & --- & $\mathcal{U}(0,1)$ & --- & Linear limb-darkening parametrization. \\
        ~~~$GP_{\sigma, \textnormal{Hazwelwood}}$ & --- & $\mathcal{J}(0.1,12000^2)$ & --- & Amplitude of GP component. \\
        ~~~$GP_{y,\textnormal{Hazwelwood}}$ & --- & $\mathcal{J}(0.01,50^2)$ & --- & Pixel y-centroid GP componenent. \\
        \hline
        Parameters for LCO z band \\
        ~~~$D_{\textnormal{LCO,z}}$ & 1.0 (fixed) & --- & --- & Dilution factor for LCO z band. \\
        ~~~$M_{\textnormal{LCO,z}}$ & $\mathcal{N}(0,0.1^2)$ & --- & ppm & Relative flux offset for LCO z band. \\
        ~~~$\sigma_{w,\textnormal{LCO,z}}$ & $\mathcal{J}(0.1,10000^2)$ & --- & ppm & Extra jitter term for LCO z band lightcurve. \\
        ~~~$q_{1,\textnormal{LCO,z}}$ & $\mathcal{U}(0,1)$ & --- & --- & Linear limb-darkening parametrization. \\
        \hline
        Parameters for LCO i band \\
        ~~~$D_{\textnormal{LCO,i}}$ & 1.0 (fixed) & 1.0 (fixed) & --- & Dilution factor for LCO i band. \\
        ~~~$M_{\textnormal{LCO,i}}$ & $\mathcal{N}(0,0.1^2)$ & $\mathcal{N}(0,0.1^2)$ & ppm & Relative flux offset for LCO i band. \\
        ~~~$\sigma_{w,\textnormal{LCO,i}}$ & $\mathcal{J}(0.1,10000^2)$ & $\mathcal{J}(0.1,5000^2)$ & ppm & Extra jitter term for LCO i band lightcurve. \\
        ~~~$q_{1,\textnormal{LCO,i}}$ & $\mathcal{U}(0,1)$ & $\mathcal{U}(0,1)$ & --- & Linear limb-darkening parametrization. \\
        ~~~$GP_{\sigma, \textnormal{LCO,i}}$ & $\mathcal{J}(0.1,10000^2)$ & $\mathcal{J}(0.1,10000^2)$ & --- & Amplitude of GP component. \\
        ~~~$GP_{t,\textnormal{LCO,i}}$ & $\mathcal{J}(0.01,10^2)$ & --- & --- & Time GP componenent. \\
        ~~~$GP_{FWHM,\textnormal{LCO,i}}$ & --- & $\mathcal{J}(0.01,100^2)$ & --- & FWHM GP componenent. \\
        ~~~$GP_{Sky\ flux,\textnormal{LCO,i}}$ & --- & $\mathcal{J}(0.01,100^2)$ & --- & Sky Flux GP componenent. \\
        \hline
        
        Parameters for El Sauce \\
        ~~~$D_{\textnormal{El\ Sauce}}$ & 1.0 (fixed) & 1.0 (fixed) & --- & Dilution factor for El Sauce. \\
        ~~~$M_{\textnormal{El\ Sauce}}$ & $\mathcal{N}(0,0.1^2)$ & $\mathcal{N}(0,0.1^2)$ & ppm & Relative flux offset for El Sauce. \\
        ~~~$\sigma_{w,\textnormal{El\ Sauce}}$ & $\mathcal{J}(0.1,10000^2)$ & $\mathcal{J}(0.1,5000^2)$ & ppm & Extra jitter term for El Sauce lightcurve. \\
        ~~~$q_{1,\textnormal{El\ Sauce}}$ & $\mathcal{U}(0,1)$ & $\mathcal{U}(0,1)$ & --- & Linear limb-darkening parametrization. \\
        ~~~$GP_{\sigma, \textnormal{El\ Sauce}}$ & $\mathcal{J}(0.1,10000^2)$ & $\mathcal{J}(0.1,150^2)$ & --- & Amplitude of GP component. \\
        ~~~$GP_{rho,\textnormal{El\ Sauce}}$ & --- & $\mathcal{J}(0.001,30^2)$ & --- & Rho for Matern GP componenent. \\
        ~~~$GP_{timescale,\textnormal{El\ Sauce}}$ & --- & $\mathcal{J}(0.001,30^2)$ & --- & Timescale for Matern GP componenent. \\
        
        \hline
        \hline
    \end{tabular}
\end{table*}

\begin{table*}
    \centering
    \caption{Continuation of Table \ref{tab:priors1}.}  
    \label{tab:priors2}
    \begin{tabular}{lcccl} 
        \hline
        \hline
        Parameter name & Prior (\tessstarnameone b) & Prior (\tessstarnametwo b) & Units & Description \\
                \hline
                \hline  

        Parameters for TRAPPIST-S \\
        ~~~$D_{\textnormal{TRAPPIST}}$ & 1.0 (fixed) & --- & --- & Dilution factor for TRAPPIST-S. \\
        ~~~$M_{\textnormal{TRAPPIST}}$ & $\mathcal{N}(0,0.1^2)$ & --- & ppm & Relative flux offset for TRAPPIST-S. \\
        ~~~$\sigma_{w,\textnormal{TRAPPIST}}$ & $\mathcal{J}(0.1,10000^2)$ & --- & ppm & Extra jitter term for TRAPPIST-S lightcurve. \\
        ~~~$q_{1,\textnormal{TRAPPIST}}$ & $\mathcal{U}(0,1)$ & --- & --- & Linear limb-darkening parametrization. \\
        ~~~$\theta_{0,\textnormal{TRAPPIST}}$ & $\mathcal{U}(-0.5,0.5)$ & --- & ppm & Offset value applied to account for meridian flip. \\
        \hline
        RV parameters\\
        ~~~$\mu_{\textnormal{FEROS}}$ & $\mathcal{N}( 5939.0783,.5^2)$ & $\mathcal{N}(21392.22,15^2)$ & m/s & Systemic velocity for FEROS. \\
        ~~~$\sigma_{w,\textnormal{FEROS}}$ & $\mathcal{J}(0.1,100^2)$ & 0.0 (fixed) & m/s & Extra jitter term for FEROS. \\
        ~~~$\mu_{\textnormal{CORALIE}}$ &$\mathcal{N}(5885.6659,15^2)$& --- & m/s & Systemic velocity for CORALIE. \\
        ~~~$\sigma_{w,\textnormal{CORALIE}}$ & 0.0 (fixed) & --- & m/s & Extra jitter term for CORALIE. \\
    
        \hline
        \hline
 .

    \end{tabular}
\end{table*}

\subsection{Joint-Analysis of \tessstarnameone} \label{subsec:toi150analysis}
As a recollection of what data was collected for \tessstarnameone, we have transit photometry (Figure \ref{fig:toi150_phot}) from \textit{TESS}, LCOGT z band (egress), LCOGT i band (egress), El Sauce (full), and TRAPPIST-S (full), as well as radial velocities (Figure \ref{fig:toi150_rvs}) from FEROS (20 points) and CORALIE (3 points). 

\subsubsection{Flux Contamination Possibility} \label{subsec:toi150_fluxcontamination}
Because \textit{TESS} has a large pixel size of 21" it is particularly important to search for nearby sources which could pollute the aperture requiring dilution factors ($D$) to be taken into account \citep[see Sections 2.1 and 3.1.2 in][]{juliet:2018}.
\tessstarnameone\ (\textit{Gaia} DR2 5262709709389254528, $G_{rp}$ magnitude of 10.85) may face some obstacles with nearby neighbors, where there are two that have relatively low magnitudes (14.20, \textit{Gaia} DR2 5262709881187945344, $\sim$41" $\approx$ 2 \textit{TESS} pixels; 11.98, \textit{Gaia} DR2 5262706681434867968, $\sim$62" $\approx$ 3 \textit{TESS} pixels), and the other nearby targets are not significantly bright enough.

Because the \textit{Gaia} $G_{rp}$-band and the \textit{TESS} band are quite similar, we can approximate what the dilution factor for \textit{TESS} ($D_{TESS}$) would be using equation 2 in \cite{juliet:2018} to get $D\approx0.71$ (assuming that the two bright objects are within the same \textit{TESS} pixel). We therefore allow the \textit{TESS} dilution factor to vary uniformly with the conservative lower bound of 0.5 to 1.0, with the idea in mind that the other targets are probably not impacting the flux significantly. Indeed, we do find that $D_{TESS}$ is consistent with 1 (0.9699; Table \ref{tab:posteriors1}). As for the other photometric instruments, the dilution factors are fixed to 1.0 as there is no indications of flux contamination.

\subsubsection{GP Hyperparameters \& Instrumental Jitter Terms}
\label{subsec:toi150_gpjitter}
The \textit{TESS} photometric data appears clean and well-behaved whereas the LCO z and i band, the El Sauce, and TRAPPIST-S data might have some dependencies on other potential factors. To see which additional factors are necessary to take into account, photometry-only fits first were made with each photometric instrument, and the posterior log-evidences were compared between fits without any detrending parameters and fits accounting for possible systematic trends using a Gaussian Process (GP) regression with a multi-dimensional squared-exponential kernel combining multiple components in time, airmass, centroid position, full-width half maximum (FWHM) and/or sky flux, if available. It was found that for the LCO z and i band photometry no additional terms are needed to correct the photometry from systematics other than a flux offset. For El Sauce photometry, we found that airmass was an important regressor to take into account with a GP. Finally, for the TRAPPIST-S photometry, we found that no additional GP was needed --- however, the meridian flip offset flux has to be modelled. For this, we simply added an extra parameter ($\theta_0$) 
that accounts for an additive flux offset at the (known) time of the meridian flip.

Aside from the GP components, we also considered possible jitter terms (i.e. values added in quadrature to the formal error bars of the data) for both the photometry and the radial-velocities. Some were found to be consistent with 0, specifically $\sigma_{w,TESS}$ and  $\sigma_{w,CORALIE}$, and therefore these parameters are set to 0 for the final fits; whereas the others ($\sigma_{w,LCO z}$, $\sigma_{w,LCO i}$, and $\sigma_{w,FEROS}$) are left to be free. 

\subsubsection{Final Model Parameters}\label{subsubsec:toi150finalparams}
With the whole set up complete, we perform two main runs for a circular and eccentric model by keeping every parameter prior identical except for $\sqrt{e_b}\sin \omega_b$ and $\sqrt{e_b}\cos \omega_b$, which were fixed for the circular model and free for the eccentric model. Interestingly enough, this hot Jupiter prefers an eccentric orbit ($e=0.26$) rather than a circular one ($\Delta$ln$Z$>80) and the posterior results can found in Tables \ref{tab:posteriors1}, \ref{tab:posteriors2}, \& \ref{tab:derivedparams} alongside the prior table for the final fit in Tables \ref{tab:priors1}, \ref{tab:priors2}. 

\subsubsection{Signals in the Residuals} \label{subsec:toi150_resid}
After performing a 1-planet model fit, the radial velocity residuals were checked for additional potential signals. By eye and by the GLS periodogram \citep{gls_zechmeister:2009}, no signals suggestive of being above the significance level of the False Alarm Probability (FAP) were seen \citep[Eq. 24 in][]{gls_zechmeister:2009}. That being said, there is \textit{some} hint of power around $\sim$ 10 days. To further investigate if it is possible that there is evidence for a 2-planet model, supplementary fits were carried out on \textit{just} the radial velocities from FEROS and CORALIE. Using wide uniform priors for the period and semi-amplitude of a second signal with \codename, we found that indeed the posterior period for an additional, non-zero amplitude signal in the data peaks at about ten days. However, when the log-evidences of the 1-planet and 2-planet models are compared, the difference is not significant ($\Delta \ln Z \lesssim 2$) and thus the simpler, 1-planet model is favored by the current data and for this reason, we do not continue to investigate the secondary signal further at this point. 

\subsection{Joint-Analysis of \tessstarnametwo} \label{subsec:toi163analysis}

For \tessstarnametwo, we have transit photometry (see Figure \ref{fig:toi163_phot}) from \textit{TESS}, CHAT (ingress), Hazelwood (ingress), LCO i band (full), and El Sauce (full), along with radial velocities from FEROS (Figure \ref{fig:toi163_rvs}). The step process for modeling fits with \tessstarnametwo\ is essentially the same as for \tessstarnameone\ with minor differences. 

\subsubsection{Flux Contamination Possibility}
Fortunately, \tessstarnametwo\ (\textit{Gaia} DR2 4651366259202463104, $G_{rp}$ magnitude of 10.82) doesn't have any neighboring \textit{Gaia} DR2 targets with a large enough flux to impact the light curve, however, there are \textit{plenty} of faint objects that \textit{might} have some influence, and therefore we let the dilution factor ($D_{TESS}$) be free for just the \textit{TESS} photometry. In fact, if we perform a rough estimation, there are about 20 objects within one \textit{TESS} pixel with magnitudes $>$18, so if we assume 20 objects with worse-case scenario magnitudes of 18, this translates to a $D$ of $\sim$0.9626. This actually corresponds quite well with the dilution value we get from the final fit of 0.96996 (see Table \ref{tab:posteriors1}). 

In addition to no bright nearby \textit{Gaia} objects, speckle data from Gemini/DSSI in both the R (692nm) and I (880nm) wavelengths show that there are no significant sources of light nearby (Figure \ref{fig:toi163_contrastcurve}). Therefore, this further confirms the planetary nature of the signal found in the light curve and radial velocities and allows us to fix the dilution factors of the other photometric instruments to 1.0.

\subsubsection{GP Hyperparameters \& Instrumental Jitter Terms} \label{subsec:toi163_gpjitter}
While the \textit{TESS} and CHAT data are relatively well-behaved, the Hazelwood, LCO i band, and El Sauce data show clear signs of systematic effects, therefore, we performed additional model fits with and without GP components, in the same manner as we explained in Section \ref{subsec:toi150_gpjitter} for \tessstarnameone. We decided to consider 1 GP component (y pixel centroid) for the  Hazelwood photometry, 2 GP components (FWHM, sky flux) for the LCO i band photometry,  and an exponential and Matern GP kernel (time) for the El Sauce photometry.

As for the jitter terms, we encounter that $\sigma_{w,TESS}$, $\sigma_{w,CHAT}$, and $\sigma_{w,FEROS}$ can be fixed to 0, whereas $\sigma_{w,Hazelwood}$, $\sigma_{w,LCO,i}$, and $\sigma_{w,El\ Sauce}$ will be allowed to be free in the fit.

\subsubsection{Final Model Parameters} \label{subsubsec:toi163finalparams}
As with \tessstarnameone, we perform circular and eccentric model fits, finding that the circular model is ever so slightly preferred ($\Delta$ln$Z$<2). The full posterior information is in Tables \ref{tab:posteriors1}, \ref{tab:posteriors2}, \& \ref{tab:derivedparams} where the priors are located in Tables \ref{tab:priors1}, \ref{tab:priors2}.

\subsubsection{Signals in the Residuals} \label{subsec:toi163_resid}
Following the same ideology as in Section \ref{subsec:toi150_resid}, we checked the radial velocity residuals for additional signals and found suggestions in the residuals for an extra periodic signal (Figure \ref{fig:toi163_rvs}). Looking at the GLS periodogram of the radial velocity residuals, a bump around 34 days is present --- it is, however, not above any significance level. 2-planet models fits on \textit{just} the radial velocities from FEROS were performed, again trying wide uniform priors in the period and semi-amplitude of a possible signal. The posterior period of this additional possible signal was 37 days --- however, the log-evidence of this 2-planet fit was also not significantly better than the 1-planet fit ($\Delta \ln Z \lesssim 2$), and thus the 1-planet fit model is preferred and the potential signal is not further explored.

\begin{table*}
    \centering
    \caption{Presented below are the posterior parameters obtained for \tessstarnameone b \tessstarnametwo b using \codename. Priors can be found in Tables \ref{tab:priors1} \& \ref{tab:priors2}. The continuation of posterior values can be found in Table \ref{tab:posteriors2}.}
    \label{tab:posteriors1}
    \begin{tabular}{lcc} 
        \hline
        \hline
        Parameter name & Posterior estimate$^a$ for \tessstarnameone b & Posterior estimate$^a$ for \tessstarnametwo b \\
        \hline
        \hline
        Posterior parameters & & \\[0.1cm]
        ~~~$P_b$ (days) &$5.857487^{+0.000089}_{-0.000097}$ & $4.231306^{+0.000063}_{-0.000057}$\\[0.1 cm]
        ~~~$t_{0,b}$ (BJD UTC) &$2458326.27730^{+0.00086}_{-0.00089}$ & $2458328.8797^{+0.00062}_{-0.00063}$ \\[0.1 cm]
        ~~~$\rho_*$ (kg/m$^3$) & $538^{+15}_{-16}$ & $459^{+24}_{-25}$\\[0.1 cm]
        ~~~$r_{1,b}$ & $0.552^{+0.077}_{-0.115}$ & $0.577^{+0.035}_{-0.038}$\\[0.1 cm]
        ~~~$r_{2,b}$ & $0.0826^{+0.0012}_{-0.0011}$ & $0.091^{+0.0016}_{-0.0015}$\\[0.1 cm]
        ~~~$K_{b}$ (m/s) & $240^{+11}_{-11}$ & $120^{+12}_{-11}$\\[0.1 cm]
        ~~~$e_b$ & $0.262^{+0.045}_{-0.037}$ & 0 (fixed$^b$, $< 0.091$)\\[0.1 cm]
        \hline
        Posterior parameters for \textit{TESS} & \\[0.1cm]
        ~~~$D_{\textnormal{TESS}}$ & $0.9959^{+0.0028}_{-0.0053}$ & $0.970^{+0.012}_{-0.030}$\\[0.1 cm]
        ~~~$M_{\textnormal{TESS}}$ (ppm) &$7^{+20}_{-20}$ & $-1^{+20}_{-21}$\\[0.1 cm]
        ~~~$\sigma_{w,\textnormal{TESS}}$ (ppm) & 0 (fixed$^b$, $< 87$) & 0 (fixed$^b$, $< 90.3$) \\[0.1 cm]
        ~~~$q_{1,\textnormal{TESS}}$ & $0.68^{+0.19}_{-0.22}$& $0.45^{+0.27}_{-0.21}$\\[0.1 cm]
        ~~~$q_{2,\textnormal{TESS}}$ & $0.076^{+0.092}_{-0.050}$ & $0.14^{+0.19}_{-0.09}$\\[0.1 cm]
        \hline
        Posterior parameters for CHAT & \\[0.1cm]
        ~~~$M_{\textnormal{CHAT}}$ (ppm) & --- &$7^{+248}_{-265}$ \\[0.1 cm]
        ~~~$\sigma_{w,\textnormal{CHAT}}$ (ppm) & --- & 0 (fixed$^b$, $< 361$)  \\[0.1 cm]
        ~~~$q_{1,\textnormal{CHAT}}$ & --- & $0.75^{+0.09}_{-0.09}$\\[0.1 cm]
        \hline
        Posterior parameters for Hazelwood & \\[0.1cm]
        ~~~$M_{\textnormal{Hazelwood}}$ (ppm) & --- &$3904^{+1917}_{-2355}$ \\[0.1 cm]
        ~~~$\sigma_{w,\textnormal{Hazelwood}}$ (ppm) & --- & $3154^{+220}_{-206}$ \\[0.1 cm]
        ~~~$q_{1,\textnormal{Hazelwood}}$ & --- & $0.54^{+0.17}_{-0.18}$\\[0.1 cm]
        ~~~$GP_{\sigma, \textnormal{Hazelwood}}$ (ppm) & --- &$3591^{+2342}_{-1259}$ \\[0.1 cm]
        ~~~$GP_{y, \textnormal{Hazelwood}}$ & ---  & $6.97^{+11.45}_{-3.65}$ \\[0.1 cm]
        \hline
        Posterior parameters for LCO z band & \\[0.1cm]
        ~~~$M_{\textnormal{LCO,z}}$ (ppm) &$-258^{+169}_{-163}$ &---\\[0.1 cm]
        ~~~$\sigma_{w,\textnormal{LCO,z}}$ (ppm) & $1096^{+110}_{-100}$ &---\\[0.1 cm]
        ~~~$q_{1,\textnormal{LCO,z}}$ & $0.404^{+0.083}_{-0.050}$&---\\[0.1 cm]
        \hline
        Posterior parameters for LCO i band & \\[0.1cm]
        ~~~$M_{\textnormal{LCO,i}}$ (ppm) &$-1317^{+172}_{-182}$ &$-7744^{+1174}_{-1145}$\\[0.1 cm]
        ~~~$\sigma_{w,\textnormal{LCO,i}}$ (ppm) & $1366^{+79}_{-74}$ & $22515^{+170}_{-144}$ \\[0.1 cm]
        ~~~$q_{1,\textnormal{LCO,i}}$ & $0.179^{+0.089}_{-0.085}$& $0.21^{+0.14}_{-0.12}$\\[0.1 cm]
        ~~~$GP_{\sigma, \textnormal{LCO,i}}$ & --- &$4725^{+762}_{-593}$ \\[0.1 cm]
        ~~~$GP_{FWHM, \textnormal{LCO,i}}$ & --- &$13.1^{+14.3}_{-6.3}$ \\[0.1 cm]
        ~~~$GP_{skyflux, \textnormal{LCO,i}}$ & --- &$27^{+34}_{-12}$ \\[0.1 cm]
        \hline
        Posterior parameters for El Sauce & \\[0.1cm]
        ~~~$M_{\textnormal{El\ Sauce}}$ (ppm) &$-6374^{+1675}_{-1392}$ &$-1772^{+18632}_{-12903}$\\[0.1 cm]
        ~~~$\sigma_{w,\textnormal{El\ Sauce}}$ (ppm) & $4449^{+242}_{-228}$ & $2457^{+144}_{-145}$\\[0.1 cm]
        ~~~$q_{1,\textnormal{El\ Sauce}}$ & $0.73^{+0.15}_{-0.19}$& $0.34^{+0.29}_{-0.22}$\\[0.1 cm]
        ~~~$GP_{\sigma,\textnormal{El\ Sauce}}$ & $3608^{+1442}_{-1021}$&$16.2^{+26.2}_{-8.9}$\\[0.1 cm]
        ~~~$GP_{airmass,\textnormal{El\ Sauce}}$ & $18^{+17}_{-14}$ & ---\\[0.1 cm]
        ~~~$GP_{rho, \textnormal{El\ Sauce}}$ & --- &$2.4^{+9.3}_{-2.0}$ \\[0.1 cm]
        ~~~$GP_{timescale, \textnormal{El\ Sauce}}$ & --- &$0.79^{+4.52}_{-0.65}$ \\[0.1 cm]
        \hline
        \hline
    \end{tabular}
    \begin{tablenotes}
      \small
      \item $^a$ Error bars denote the $68\%$ posterior credibility intervals.
      \item $^b$ Upper limits denote the 95\% upper credibility interval of fits.
    \end{tablenotes}
\end{table*}

\begin{table*}
    \centering
    \caption{Continuation for Table \ref{tab:posteriors1}.}
    \label{tab:posteriors2}
    \begin{tabular}{lcc} 
        \hline
        \hline
        Parameter name & Posterior estimate$^a$ for \tessstarnameone b & Posterior estimate$^a$ for \tessstarnametwo b \\
        \hline
        \hline
        Posterior parameters for TRAPPIST-S & \\[0.1cm]
        ~~~$M_{\textnormal{TRAPPIST-S}}$ (ppm) &$-6673^{+207}_{-215}$ & ---\\[0.1 cm]
        ~~~$\sigma_{w,\textnormal{TRAPPIST-S}}$ (ppm) & $4122^{+99}_{-95}$& ---\\[0.1 cm]
        ~~~$q_{1,\textnormal{TRAPPIST-S}}$ & $0.54^{+0.12}_{-0.12}$& ---\\[0.1 cm]
        ~~~$\theta_{0,\textnormal{TRAPPIST-S}}$ & $-0.00500^{+0.00041}_{-0.00043}$& ---\\[0.1 cm]
        \hline
        Posterior RV parameters & \\[0.1cm]
        ~~~$\mu_{\textnormal{FEROS}}$ (m/s) & $5939.0^{+7.3}_{-7.2}$ & $21393.7^{+6.7}_{-6.6}$\\[0.1 cm]
        ~~~$\sigma_{w,\textnormal{FEROS}}$ (m/s) & $32.8^{+7.9}_{-6.6}$ & 0 (fixed$^b$, $< 43$)\\[0.1 cm]
        ~~~$\mu_{\textnormal{CORALIE}}$ (m/s) & $5887^{+12}_{-13}$ & ---\\[0.1 cm]
        ~~~$\sigma_{w,\textnormal{CORALIE}}$ (m/s) & 0 (fixed$^b$, $< 36 $) & ---\\[0.1cm]
        \hline
        \hline
    \end{tabular}
    \begin{tablenotes}
      \small
      \item $^a$ Error bars denote the $68\%$ posterior credibility intervals.
      \item $^b$ Upper limits denote the 95\% upper credibility interval of fits.
    \end{tablenotes}
\end{table*}

\begin{table*}
    \centering
    \caption{Presented below are the derived planetary parameters obtained for \tessstarnameone b and \tessstarnametwo b using the posterior values from Tables \ref{tab:posteriors1} \& \ref{tab:posteriors2}.}
    \label{tab:derivedparams}
    \begin{tabular}{lcc} 
        \hline
        \hline
        Parameter name & Posterior estimate$^a$ for \tessstarnameone b & Posterior estimate$^a$ for \tessstarnametwo b \\
        \hline
        \hline
        Derived transit parameters for & & \\[0.1cm]
        ~~~$R_p/R_*$ & $0.0826^{+0.0012}_{-0.0011}$ & $0.09082^{+0.0016}_{-0.0015}$\\[0.1 cm]
        ~~~$b = (a/R_*)\cos(i_p)$ & $0.33^{+0.12}_{-0.17}$ & $0.365^{+0.053}_{-0.057}$ \\[0.1 cm]
        ~~~$a_{b}/R_*$ & $9.917^{+0.092}_{-0.099}$ & $7.57^{+0.13}_{-0.14}$\\[0.1 cm]
        ~~~$i_p$ (deg) & $88.09^{+0.98}_{-0.68}$ & $87.24^{+0.47}_{-0.45}$\\[0.1 cm]
        ~~~$u_1$  & $0.124^{+0.131}_{-0.082}$ & $0.19^{+0.16}_{-0.12}$\\[0.1 cm]
        ~~~$u_2$  & $0.69^{+0.15}_{-0.21}$ & $0.48^{+0.25}_{-0.32}$ \\[0.1 cm]
        ~~~$t_T$ (hours) & $5.12^{+0.21}_{-0.18}$ & $4.93^{+0.17}_{-0.15}$ \\[0.1 cm]
        \hline
        Derived physical parameters & &\\[0.1cm]
        ~~~$M_{p}$ ($M_J$) & $2.51^{+0.12}_{-0.12}$ & $1.22^{+0.12}_{-0.12}$\\[0.1 cm]
        ~~~$R_{p}$ ($R_J$) & $1.255^{+0.021}_{-0.019}$ & $1.489^{+0.034}_{-0.034}$\\[0.1 cm]
        ~~~$\rho_{p}$ (g cm$^{-3}$) & $1.68^{+0.12}_{-0.12}$ & $0.49^{+0.059}_{-0.055}$\\[0.1 cm]
        ~~~$g_{p}$ (m s$^{-2}$) & $41.3^{+2.5}_{-2.4}$ & $14.2^{+1.5}_{-1.5}$  \\[0.1 cm]
        ~~~$a$ (AU) & $0.07037^{+0.00087}_{-0.00088}$ & $0.0580^{+0.0014}_{-0.0014}$\\[0.1 cm]
        ~~~$T_\textnormal{eq}$ (K)$^c$ & $1404.5^{+7.1}_{-6.5}$ & $1669^{+16}_{-14}$ \\[0.1 cm]
        \hline
        \hline
    \end{tabular}
    \begin{tablenotes}
      \small
      \item $^a$ Error bars denote the $68\%$ posterior credibility intervals.
      \item $^c$ Equilibrium temperatures calculated assuming 0 Bond Albedo.
    \end{tablenotes}
\end{table*}

\subsection{Stellar Density Prior} \label{sec:stellardensitycomp}
We also experimented with the impact that a stellar density prior\footnote{When $\rho_*$ is given as a prior, then $a$, the scaled semi-major axis, is no longer a model parameter.}, $\rho_*$, may have on eccentricity as well as on the stellar density itself by allowing the stellar density prior to be an uninformative Jeffrey's prior, $\mathcal{J}(1,10000)$, rather than a normal prior (as provided by Table \ref{tab:priors1}). Focusing just on the \tessstarnameone\ data since this target has a planet with eccentric behavior, we found that the eccentricities agree with each other regardless of whether $\rho_*$ was given as a normal ($e=0.26\pm 0.04$) or Jeffrey's prior ($e=0.27\pm 0.05$). Both obtained stellar densities from the eccentric fits agreed with the expected density where the distribution was accurate ($\rho_*=537^{+15}_{-16}$) but much more uncertain when $\rho_*$ was given as a Jeffrey's prior ($\rho_* = 523^{+129}_{-120}$). As for the circular fits, both obtained density distributions deviated from the expected value yet showed narrow precision; when $\rho_*$ was given as a normal prior, the deviation was mild ($\rho_* = 451^{+10}_{-11}$), where the deviation was huge for when $\rho_*$ was given as a Jeffrey's prior ($\rho_* = 25^{+3}_{-4}$). This demonstrated disagreement of stellar density distributions among the circular fits is due to the fact that the evidence for \tessstarnameone\ favors a non-circular model over a circular model. 

\subsection{Search for secondary eclipses}
A search for secondary eclipses was performed on the \textit{TESS} photometry. The expected secondary eclipse depth, assuming reflected light is the main component (i.e., a depth equal to $A_g(a/R_p)^2$, where $A_g$ is the geometric albedo) was smaller than 69$\pm$2 ppm for \tessstarnameone b, and 144$\pm$7 ppm for \tessstarnametwo b \citep[assuming $A_g<1$, which seems to be the case for hot Jupiters; see, e.g., ][]{HD:2013}. Given the \textit{TESS} data as of now solely from Sector 1, there is no significant dip at the anticipated times. They might be detectable, however, once data from future sectors is released --- see Section \ref{subsec:trans_rm} for a more in-depth discussion. Detecting phase variations \citep[as described in][]{shporer_phasevariations:2017} with the current data is not possible given the large amount of systematics present. 

\section{Discussion} \label{sec:discussion}
\begin{figure*}
	\includegraphics[height=0.65\columnwidth]{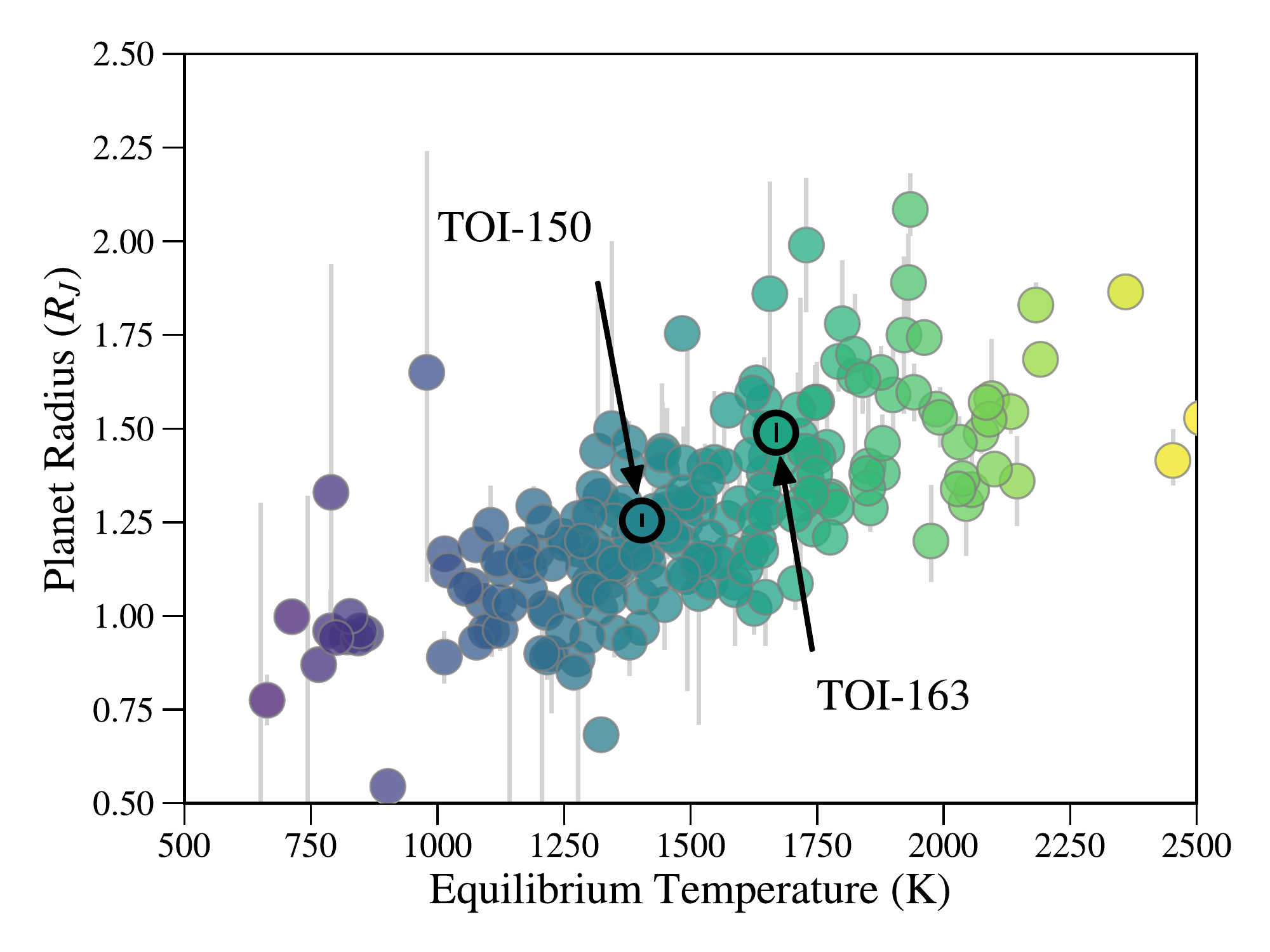}
	\includegraphics[height=0.65\columnwidth]{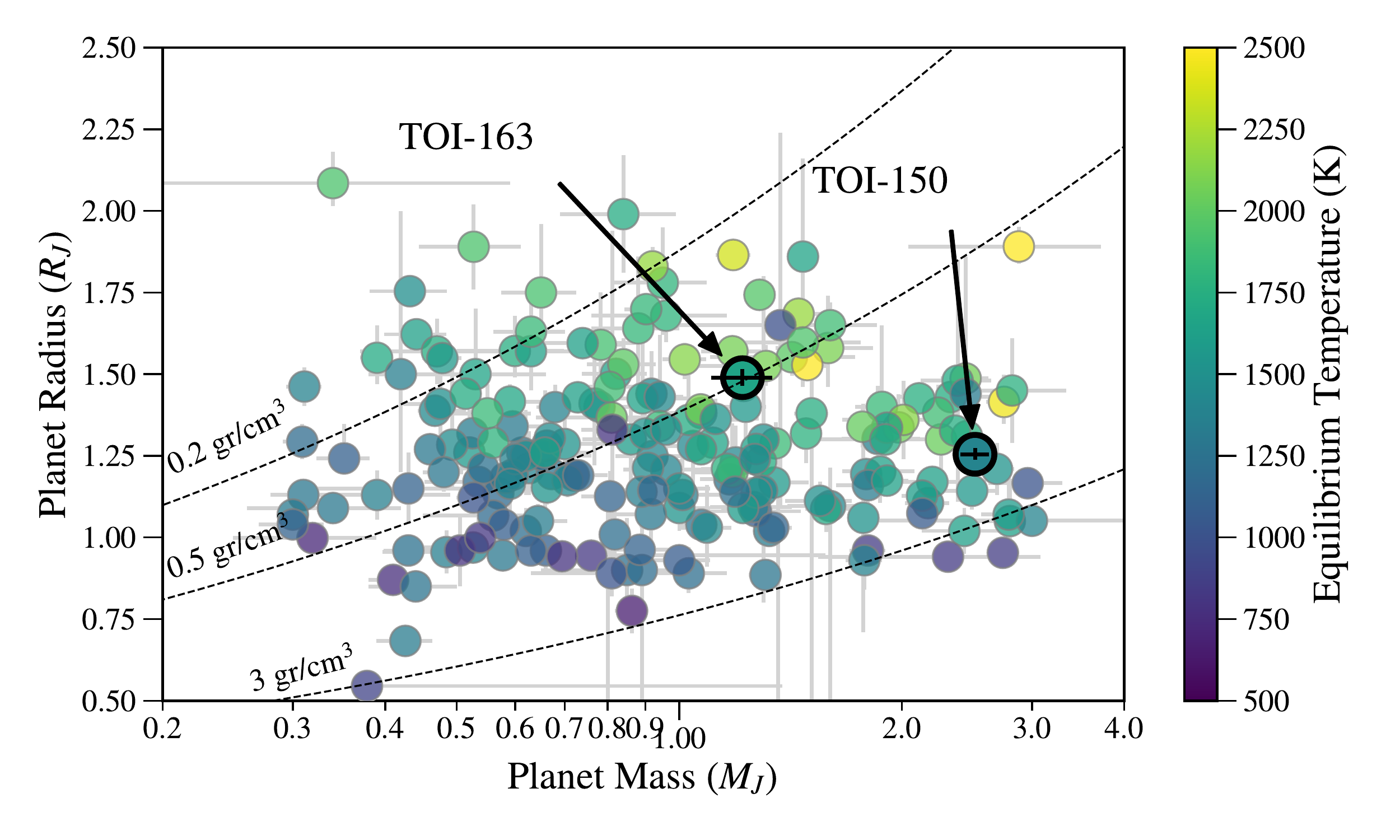}
    \caption{Radius versus equilibrium temperature (left) and a radius versus mass (right) plot of the known hot Jupiters ($0.7\leq P$ (days) $\leq10$, $0.3\leq M_p\ (M_J) \leq 3.0$) where \tessstarnameone b and \tessstarnametwo b are annotated and their error bars are plotted on top}. Note the small error for the targets characterized in this work in comparison with previously characterized systems.
    \label{fig:mrd}
\end{figure*}

\subsection{The two systems}
With the help of multiple photometric and spectroscopic instruments (which highlights the enormous contribution that a program such as TFOP can deliver to exoplanetary science) we were able to obtain tight constraints on the period and time of periastron, and thanks to precise parallax measurements from \textit{Gaia} we constrain the stellar radius, and therefore the planetary radius and semimajor-axis very well, in comparison to other known hot Jupiters\footnote{using the NASA Exoplanet Archive; exoplanetarchive.ipac.caltech.edu, accessed on 11 March 2019} (Figure \ref{fig:mrd}). \tessstarnameone b is a $1.254\pm0.016 \textnormal{R}_\textnormal{J}$ massive ($2.61^{+0.19}_{-0.12}\ \textnormal{M}_\textnormal{J}$) hot Jupiter in a $5.857$-day orbit with a peculiarly high eccentricity ($e=0.262^{+0.045}_{-0.037}$) --- discussed more in Section \ref{subsubsec:eccentricity} --- and density larger than Jupiter's ($\rho_{p}$ = $1.7\pm0.1$ g cm$^{-3}$). On the other hand, \tessstarnametwo b is an inflated hot Jupiter ($R_\textnormal{P}$ = $1.478^{+0.022}_{-0.029} R_\textnormal{J}$, $M_\textnormal{P}$ = $1.22\pm0.11 \textnormal{M}_\textnormal{J}$) on a $P$ = $4.231$-day circular orbit, with a density less than that of Saturn ($\rho_{p}$ = $0.49\pm0.05$ g cm$^{-3}$). Though \tessstarnametwo b does not appear to be an outlier in Figure \ref{fig:mrd} relative to the other planets, targets of such equilibrium temperatures are not expected to have such high radii, but rather radii of $1 M_J$ \citep{sestovic:2018}.
These two targets are quite exciting given that both of them should be observed in at least 12 sectors with \textit{TESS}. Moreover, \tessstarnameone\ and \tessstarnametwo\ are only 10.4$^\circ$ and 6.4$^\circ$, respectively, away from the center of the Continuous Viewing Zone (CVZ)\footnote{https://jwst-docs.stsci.edu/display/JTI/\\JWST+Observatory+Coordinate+System+and+Field+of+Regard} of the future James Webb Space Telescope \citep[JWST; ][]{garder_jwst:2009}. Note that the CVZ has a relatively small radius of 5$^\circ$, meaning that \tessstarnametwo\ is sitting right on the edge. In fact, both targets should be observable for more than $\sim$200 days\footnote{Figure 2, https://jwst-docs.stsci.edu/display/JTI/\\JWST+Target+Viewing+Constraints} with this future exciting space-based observatory. Though both targets are not particularly suitable for transmission spectroscopy with JWST, they both show promise for secondary eclipse observations  --- further discussed in Section \ref{subsec:trans_rm}. Both targets are moreover ideal for the Rossiter-McLaughlin (RM) effect, where an ample number of observations during the transits could be taken, allowing us to resolve the effect well and thus, gain a better grasp for the spin-orbit alignment of the system --- also explained more in Section \ref{subsec:trans_rm}.

\subsubsection{Eccentricity of \tessstarnameone b}\label{subsubsec:eccentricity}
\begin{figure}
    \includegraphics[width=\linewidth]{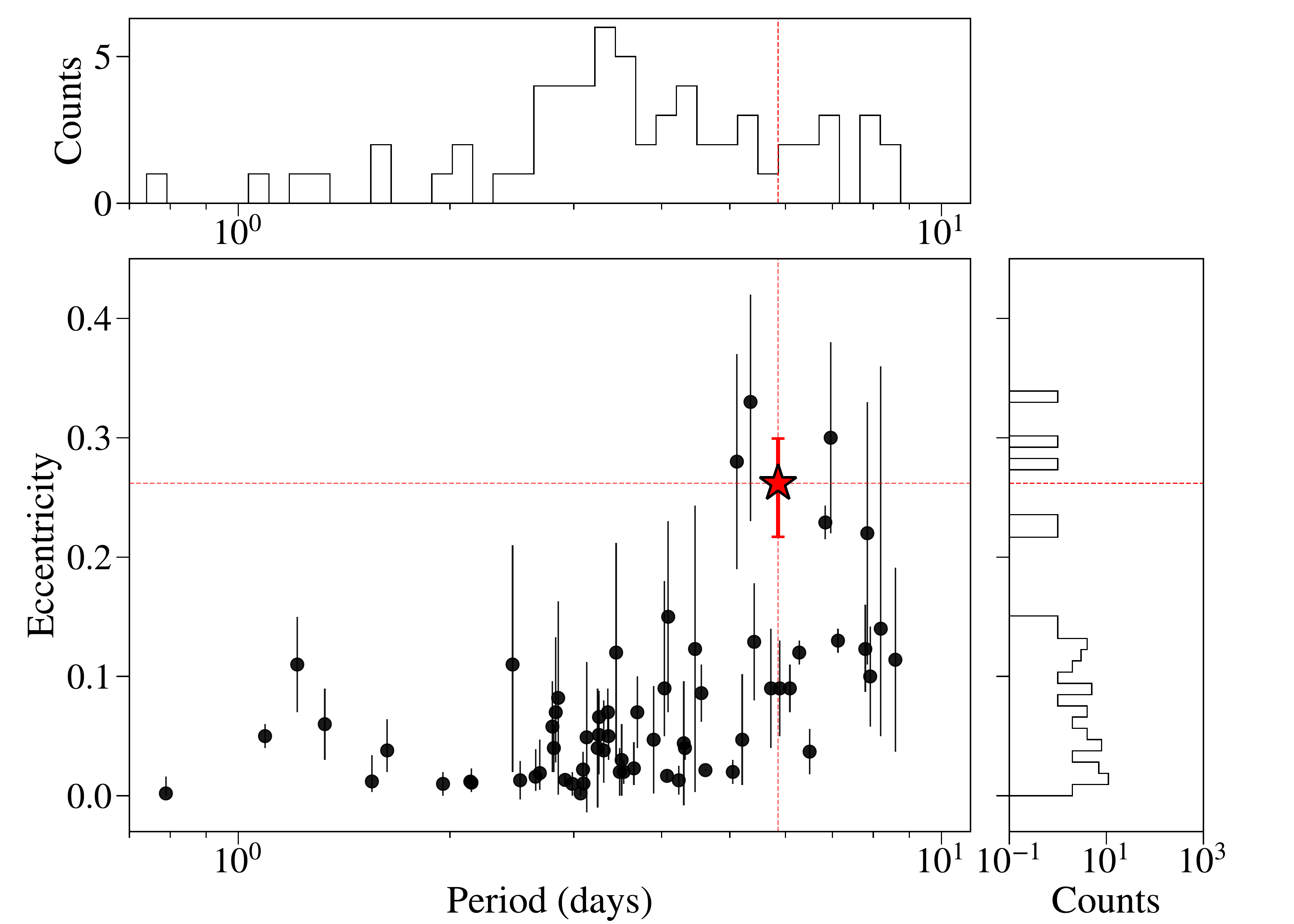}
    \caption{Eccentricities as a function of planetary period for known hot Jupiters ($0.7\leq P$ (days) $\leq10$, $0.3\leq M_p\ (M_J) \leq2.0$) where \tessstarnameone b is denoted as a red star. There are a total of 63 planets with non-zero eccentricity. Note that non-zero eccentricity planets without proper error bars were ignored given that the provided eccentricity values most likely were representing the upper eccentricity value rather than the true eccentricity.}
    \label{fig:ecc_vs_period}
\end{figure}

When we look at all the known hot Jupiters and their eccentricities (Figure \ref{fig:ecc_vs_period}), we notice that most of them have zero eccentricities. For a hot Jupiter to have a non-zero eccentricity, it either has to be currently migrating towards a circular orbit through tidal decay or it has to be excited into an eccentric orbit by, e.g., a stellar or planetary companion. For this reason, exoplanets with higher eccentricities are intriguing to follow and explore --- \tessstarnameone b is alluring for this reason.

We calculate the circularization time-scale \citep[Eq. 2 in][]{adams:2003} to be $3.46 \pm 0.68$ Gyr using a Q-factor of $10^6$ \citep{qfactor}, or 2 magnitudes larger if we adopt a Q-factor of $\sim 10^8$ \citep{Cameron:2018} since the time-scale scales linearly with Q. This time-scale is then on the same order of magnitude as the age of the star or larger ($\gtrsim2.46$ Gyr, Table \ref{tab:stellarparams}). If the time-scale were shorter than the age of the star, then we would expect to see an already circular orbit, unless there were other companions involved that could have excited the planet into an eccentric orbit. Our calculation serves as just a rough order of magnitude estimate, as the Q-factor is not so well defined for F-type stars --- work similar to that of \cite{penev:2016,Hoyer:2016_1,Hoyer:2016_2} have constrained this factor for solar-type stars to be 6.5$-$7 using massive ultra-short period giant exoplanets. Such a study is needed for F-type stars to understand whether our selected Q-factor is reasonable and, thus, if the observed circularization time-scale truly agrees with our estimated age of the system. 

\subsection{Candidates for Secondary Eclipses, Spectroscopic Transmission, and RM effect} \label{subsec:trans_rm}
As mentioned before, both targets are very close to the JWST CVZ, particularly \tessstarnametwo\ being just on the edge. This makes these targets interesting in their own right as scheduling for these targets would be easier, which would open the window for several exciting possibilities of atmospheric characterization. For transmission spectroscopy in particular, \tessstarnametwo b is a decent target \citep[with an expected atmospheric signal in transmission of $\sim$70 ppm, assuming one scale-height of variation; see, e.g., ][]{wakeford:2019} whereas \tessstarnameone b is not particularly good since the expected atmospheric signal in transmission ($\sim$20 ppm) is just hitting the noise floor of 20 ppm for JWST \citep{greene:2016}. 

\begin{figure}
    \includegraphics[width=\linewidth]{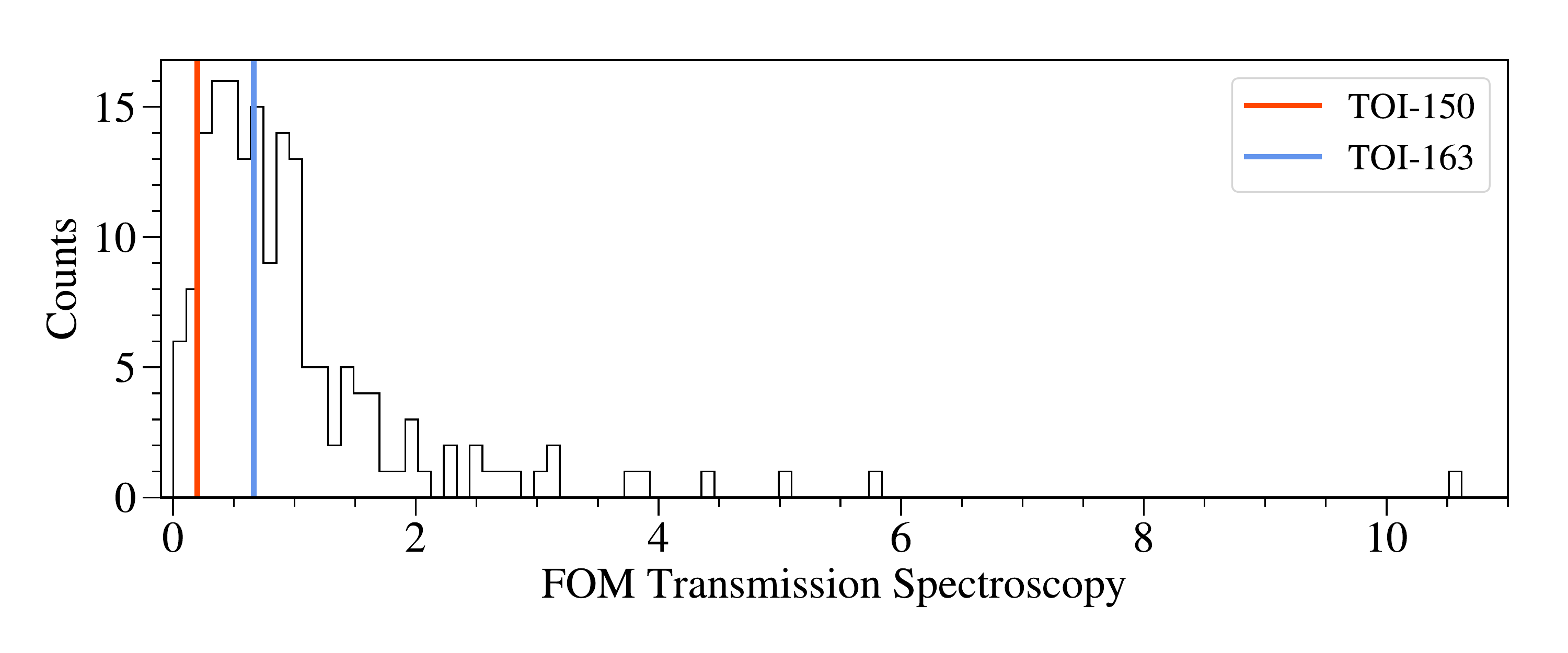}
    \includegraphics[width=\linewidth]{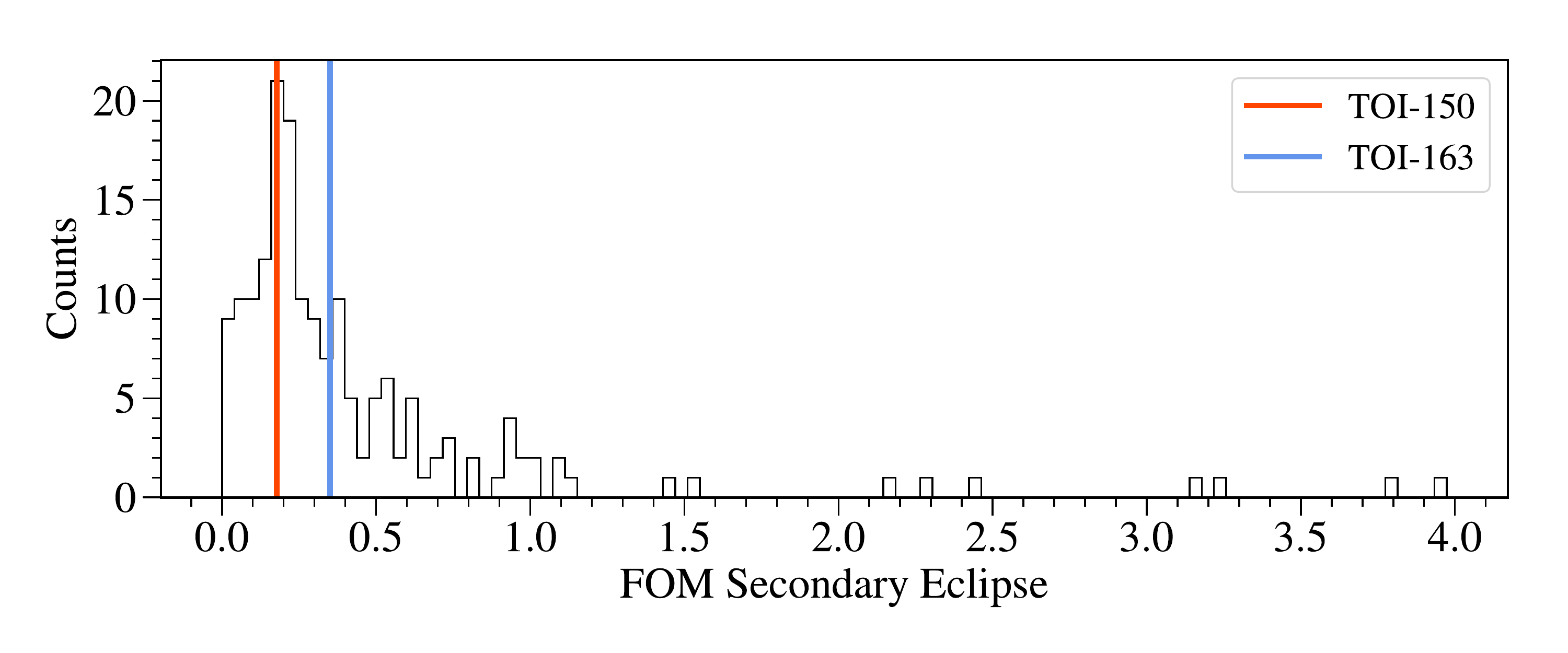}
    \caption{Histograms of the figure of merit (FOM) for both transmission spectroscopy (top) and secondary eclipses (bottom) for all known transiting hot Jupiters ($0.7\leq P$ (days) $\leq10$, $0.3\leq M_p\ (M_J) \leq2.0$) are shown. The two targets are not the top candidates for transmission spectroscopy with JWST, but will be good follow-up candidates for secondary eclipses. Note that those with all required parameters in  calculating the FOM were kept (171 in total).}
    \label{fig:fom}
\end{figure}

In general, the expected atmospheric signal alone does not tell us how good actual observations with observatories like JWST will be for the targets, as this has to be weighted against, e.g., the apparent magnitude of the targets. We thus use the figure 
of merit (FOM) introduced by \cite{zellem_fom_transspec:2017} in order to calculate how good our targets are for transmission spectroscopy studies and compare this to the known hot Jupiters. This FOM is given by 
\begin{equation*}
\textnormal{FOM}_{transspec} = \frac{\delta_{transpec}}{10^{0.2H-mag}},
\end{equation*}
where
\begin{equation*}
\delta_{transpec} = \frac{2 R_p H}{R_*^2}.
\end{equation*}
Here, $R_p$ is the planetary radius, $R_*$ is the stellar radius, and $H = k_b T_p / m g_p$ is the planetary scale-height. For calculating the scale height, the different parameters include the Boltzmann's constant, $k_b$, the planetary equilibrium temperature, $T_p$, the mean mass, $m$, that makes up the planet's atmosphere (assumed $2.3 m_{proton}$ for a hot Jupiter resembling a composition consisting mostly of $H_2$), and the gravity on the planet's surface, $g_p$. $H-mag$ in the FOM is the magnitude of the host star in the $H$ band. We present the FOM for transmission spectroscopy for all known transiting hot Jupiters in the top panel of \ref{fig:fom}. As can be seen, \tessstarnametwo\ is the best of the two here presented exoplanets for transmission spectroscopy, but it has a rather average FOM in comparison to other known hot Jupiters (Figure \ref{fig:fom}).

We repeat this exercise for our targets, but now 
for secondary eclipses following the FOM introduced in \cite{zellem_fom_eclipse:2018}, which is given by
\begin{equation*}
    \textnormal{FOM}_{eclipse} = \frac{F_p R_p^{2} F_*^{-1} R_*^{-1}}{10^{0.2H-mag}} 
\end{equation*}
where $F$ is the flux of either the planet or star and which here, for simplicity, we approximate with blackbody radiation. We find that the secondary eclipses of both targets should be observed with JWST (Figure \ref{fig:fom}). 

Both targets are deemed as highly suitable targets for follow-up Rossiter-McLaughlin (RM) observations, which can aid in determining the spin-orbit alignment between the hot Jupiter and the host star. Many hot Jupiters have been found to have large misalignments \citep{crida:2014} and the degree of misalignment can help in distinguishing between different migration theories. In addition, both targets lie just above the cutoff ($T_{eff}=6090^{+150}_{-110}$K) where we expect to see co-planar and misaligned planets \citep{triaud:2018}, which is even more so intriguing for \tessstarnameone b given its eccentric nature. Using equation 6 of \cite{gaudiwinn_rm:2007}, 

\begin{equation*}
    K_R = 52.8 \text{ m s}^{-1} \left(\frac{V_S \sin I_S}{5 \text{ km s}^{-1}} \right) \left( \frac{r}{R_J}\right)^2 \left( \frac{R}{R_\odot}\right)^{-2}
\end{equation*}

where $V_S \sin I_S$ is 7.96 and 14.08 km s$^{-1}$ for \tessstarnameone\ and \tessstarnametwo\ (Table \ref{tab:stellarparams}), respectively, $r$ is the radius of the planet, and $R$ is the radius of the star; we obtain $K_R$ values of $56.9^{+2.6}_{-2.5}$ and $121.4^{+4.9}_{-4.7}$ ms$^{-1}$ for \tessstarnameone b and \tessstarnametwo b, respectively. Given that the average spectrum exposure time is roughly 400$\sim$600 seconds with an average uncertainty of 15ms$^{-1}$ (for an instrument like FEROS) and that the transit duration is $5.12^{+0.21}_{-0.18}$ and $4.93^{+0.17}_{-0.15}$ hours for \tessstarnameone b and \tessstarnametwo b, respectively, then we would be able to obtain at least 30 and 29 observations during the transit, which is more than adequate to resolve the RM effect, making both targets optimal for these observations. 

\section{Summary}
In this paper, we have presented the 30-minute cadence \textit{TESS} discovery of two hot Jupiters, \tessstarnameone b and \tessstarnametwo b, supported by follow-up photometric and spectroscopic measurements, in which a joint fit of the transit photometry and radial velocity data was performed using the new tool \codename\ in order to thoroughly constrain the planet parameters with truly high precision. The radial velocity and speckle imaging all favor and provide evidence of the planetary nature of these detected signals. Both targets exhibit promising outcomes for investigating spin-orbit alignment using the RM effect and they both will serve as great secondary eclipse candidates considering they are very close to the JWST CVZ. \tessstarnameone b is on its own an appealing exoplanet to investigate given its high, non-zero eccentricity of 0.26, a very uncommon value among already known hot Jupiters.

\section*{Acknowledgements}
Funding for the TESS mission is provided by NASA's Science Mission directorate. We acknowledge the use of TESS Alert data, which is currently in a beta test phase, from pipelines at the TESS Science Office and at the TESS Science Processing Operations Center. This research has made use of the Exoplanet Follow-up Observation Program website as well as the Exoplanet archive, which is operated by the California Institute of Technology, under contract with the National Aeronautics and Space Administration under the Exoplanet Exploration Program. This paper includes data collected by the TESS mission, which are publicly available from the Mikulski Archive for Space Telescopes (MAST). 
D.K.\ would like to acknowledge the support from the Deutsche Forschungsgemeinschaft for the Research Unit FOR2544 "Blue Planets around Red Stars".
N.E.\ would like to thank the Gruber Foundation for its generous support to this research. 
R.B., A.J., and F.R\ acknowledge support from the Ministry for the Economy, Development, and Tourism's Programa Iniciativa Cient\'{i}fica Milenio through grant IC\,120009, awarded to the Millennium Institute of Astrophysics (MAS).
R.B.\ acknowledges additional support from FONDECYT Postdoctoral Fellowship Project 3180246.
A.J.\ acknowledges additional support from FONDECYT project 1171208 and CONICYT project BASAL AFB-170002.
TRAPPIST-South is funded by the Belgian Fund for Scientific Research (Fond National de la Recherche Scientifique, FNRS) under the grant FRFC 2.5.594.09.F, with the participation of the Swiss National Science Fundation (SNF). M.G. and E.J. are FNRS Senior Research Associates.
Based on observations made with DSSI and obtained at the Gemini Observatory, which is operated by the Association of Universities for Research in Astronomy, Inc., under a cooperative agreement with the NSF on behalf of the Gemini partnership: the National Science Foundation (United States), National Research Council (Canada), CONICYT (Chile), Ministerio de Ciencia, Tecnolog\'{i}a e Innovaci\'{o}n Productiva (Argentina), Minist\'{e}rio da Ci\^{e}ncia, Tecnologia e Inova\c{c}\~{a}o (Brazil), and Korea Astronomy and Space Science Institute (Republic of Korea). We would also like to personally thank Jen Winters, Dan Nusdeo, and Zack Hartman for taking the DSSI observations.
I.J.M.C. acknowledges support from the NSF through grant AST-1824644.
This publication was made possible through the support of a grant from the John Templeton Foundation. The opinions expressed in this publication are those of the authors and do not necessarily reflect the views of the John Templeton Foundation.





\bibliographystyle{mnras}



\appendix
\section{Author Affiliations}\label{affiliations}
$^{1}$Max-Planck-Institut f\"{u}r Astronomie, K\"{o}nigstuhl  17, 69117 Heidelberg, Germany\\
$^{2}$Center of Astro-Engineering UC, Pontificia Universidad Cat\'{o}lica de Chile, Av. Vicu\~{n}a Mackenna 4860, 782-0436 Macul, Santiago, Chile\\
$^{3}$Instituto de Astrof\'isica, Facultad de F\'isica, Pontificia Universidad Cat\'olica de Chile, Av. Vicu\~na Mackenna 4860, 782-0436 Macul, Santiago, Chile\\
$^{4}$Millennium Institute of Astrophysics, Av. Vicu\~na Mackenna 4860, 782-0436 Macul, Santiago, Chile\\
$^{5}$Space Sciences, Technologies and Astrophysics Research (STAR) Institute, Universit\'e de Li\`ege, 19C All\'ee du 6 Ao\^ut, 4000 Li\`ege, Belgium\\
$^{6}$Astrobiology Research Unit, Universit\'e de Li\`ege, 19C All\'ee du 6 Ao\^ut, 4000 Li\`ege, Belgium\\
$^{7}$Oukaimeden Observatory, High Energy Physics and Astrophysics Laboratory, Cadi Ayyad University, Marrakech, Morocco\\
$^{8}$Department of Physics, Massachusetts Institute of Technology, 182 Memorial Dr, Cambridge, MA 02142, USA\\
$^{9}$Department of Physics, Southern Connecticut State University, 501 Crescent Street, New Haven, CT, 06515, USA\\
$^{10}$Caltech/IPAC-NASA Exoplanet Science Institute, M/S 100-22, 770 S. Wilson Ave, Pasadena, CA 91106 USA\\
$^{11}$Department of Astronomy and Astrophysics, University of California, Santa Cruz, CA 95064, USA\\
$^{12}$NASA Ames Research Center, Moffett Field, CA 94035, USA\\
$^{13}$Exoplanets and Stellar Astrophysics Laboratory, Code 667, NASA Goddard Space Flight Center, Greenbelt, MD USA\\
$^{14}$Department of Physics and Kavli Institute for Astrophysics and Space Research, Massachusetts Institute of Technology, Cambridge, MA 02139, USA\\
$^{15}$Department of Earth, Atmospheric and Planetary Sciences, Massachusetts Institute of Technology, Cambridge, MA 02139, USA\\
$^{16}$Department of Aeronautics and Astronautics, MIT, 77 Massachusetts Avenue, Cambridge, MA 02139, USA\\
$^{17}$Department of Astrophysical Sciences, Princeton University, 4 Ivy Lane, Princeton, NJ 08544 USA\\
$^{18}$SETI Institute/NASA Ames Research Center\\
$^{19}$Leidos, Inc./NASA Ames Research Center\\
$^{20}$Harvard-Smithsonian Center for Astrophysics, 60 Garden St., Cambridge, MA, 02138, USA\\
$^{21}$CRESST II, NASA Goddard Space Flight Center, Greenbelt, MD, 20771, USA\\
$^{22}$Department of Astronomy, University of Maryland, College Park, MD, 20742, USA\\
$^{23}$Observatoire de l'Universit\'{e} de Gen\`{e}ve, 51 chemin des Maillettes, 1290 Versoix, Switzerland\\
$^{24}$Department of Physics and Astronomy, Vanderbilt University, Nashville, TN 37235, USA\\
$^{25}$El Sauce Observatory, Coquimbo Province, Chile\\
$^{26}$Department of Physics \& Astronomy, Swarthmore College, Swarthmore PA 19081, USA\\
$^{27}$Las Campanas Observatory, Carnegie Institution of Washington, Colina el Pino, Casilla 601 La Serena, Chile\\
$^{28}$Hazelwood Observatory, Australia\\

\section{Extra material}

\begin{table*}
 \centering 
 \begin{center}
\caption{RV data for \tessstarnameone\ and \tessstarnametwo. Data will be available online in machine-readible format.}
 \label{tab:rvdata}
 \begin{threeparttable}
  \centering
  \begin{tabular}{ lccccr }
   \hline
   \hline
     BJD &  RV (m/s) &  $\sigma_{RV}$ (m/s) & BIS (m/s) & $\sigma_{BIS}$ (m/s) & Instrument\\
    \hline
    \hline
    \tessstarnameone &&& \\
    \hline
    2458380.90067285 & 5759.5 & 20.3 & -170.0 & 16.0 & FEROS \\ 
    2458382.88380768 & 6173.3 & 18.1 & 24.0 & 15.0 & FEROS \\ 
    2458383.87194905 & 6111.3 & 17.9 & 39.0 & 14.0 & FEROS \\ 
    2458404.88308931 & 5742.8 & 21.1 & 45.0 & 16.0 & FEROS \\ 
    2458405.88147085 & 6123.2 & 41.5 & -55.0 & 28.0 & FEROS \\ 
    \vdots & \vdots & \vdots & \vdots & \vdots& \vdots\\
    \hline
    \tessstarnametwo &&& \\
    \hline
    
    2458378.85013241 & 21568.5 & 40.7 & 153.0 & 15.0 & FEROS \\ 
    2458380.89084693 & 21207.4 & 40.1 & 142.0 & 14.0 & FEROS \\ 
    2458382.87693218 & 21457.0 & 36.6 & 66.0 & 13.0 & FEROS \\ 
    2458404.85311362 & 21539.5 & 48.6 & 24.0 & 17.0 & FEROS \\ 
    2458406.82378293 & 21393.1 & 49.6 & 91.0 & 17.0 & FEROS \\
    \vdots & \vdots & \vdots & \vdots & \vdots& \vdots\\
   
   \hline
   \hline
   \end{tabular}
      
 \end{threeparttable}
 \end{center}
 \end{table*}


\bsp	
\label{lastpage}

\end{document}